\documentclass{article}
\usepackage{amsmath,amssymb}
\usepackage{graphicx}
\usepackage{color}
\usepackage{bm}% bold math
\newcommand{\nn}{\nonumber}

\newcommand{\be}{\begin{eqnarray}}
\newcommand{\ee}{\end{eqnarray}}
\newcommand{\Slash}[1]{{\ooalign{\hfil/\hfil\crcr$#1$}}}
\newcommand{\ul}{\underline}
\newcommand{\wt}{\widetilde}
\newcommand{\ve}{\varepsilon}
\newcommand{\vs}{\varsigma}
\newcommand{\vt}{\vartheta}

\newcommand{\mf}{\mathfrak}

\begin{document}

%\begin{flushright} 
%%LAPTH-***/17 \\ % \\ KEK-Cosmo-, 
%%KEK-TH-****
%\end{flushright}
%\vspace*{5mm}

\begin{center}
{\bf
%\begin{spacing}{1.5}
{\LARGE Asymptotic Scale Invariance\\ and its Consequences}\\
%\vspace{4pt}
%{\Large Title}
%\end{spacing}
}
\vspace*{7mm}
{\large
Mikhail Shaposhnikov and Kengo Shimada
}
%{\large
%Satoshi Iso${}^{\; a,b}$, Pasquale D. Serpico${}^{\; c}$ and Kengo Shimada${}^{\; c}$
%}
%{\large
%Satoshi Iso${}^{\; a,b}$, Kazunori Kohri${}^{\; a,b}$ and Kengo Shimada${}^{\; c}$
%}
\vspace{3mm}

{\it Institute of Physics, Laboratory for Particle Physics and Cosmology (LPPC),\\ \'Ecole Polytechnique F\'ed\'erale de Lausanne (EPFL),\\ CH-1015 Lausanne, Switzerland  }
%{\sl\small  LAPTh, Universit\'e de Savoie, CNRS, B.P. 110, F-74941 Annecy-le-Vieux, France \\ }

%\vspace{8pt}
%e-mails: {\small \it satoshi.iso(at)kek.jp, kohri(at)post.kek.jp , kengo.shimada(at)lapth.cnrs.fr }
%\vspace{8pt}

\end{center}

\vspace{0mm}

\begin{abstract} 
Scale invariance supplemented by the requirement of the absence of new heavy particles may play an important role in addressing the hierarchy problem. We discuss how the Standard Model may become scale-invariant at the quantum level above a certain value of the Higgs field value without addition of new degrees of freedom and analyze phenomenological and cosmological consequences of this setup, in particular, possible metastability of the electroweak vacuum and Higgs inflation.
\end{abstract}

\section{Introduction \label{Introduction}}

The discovery of the Higgs boson at the LHC \cite{Aad:2012tfa,Chatrchyan:2012xdj} completed the Standard Model (SM) of particle physics. The observed Higgs mass $\simeq 125$ GeV tells us that the SM can be valid up to very high energy scale. More explicitly, the Higgs mass falls within the range where the Landau pole in the Higgs self-coupling, as well as in the SM $U(1)$ gauge coupling, is absent below the Planck scale \cite{Maiani:1977cg,Cabibbo:1979ay,Lindner:1985uk} and the longevity of the electroweak (EW) vacuum is assured \cite{Bezrukov:2012sa,Degrassi:2012ry,Buttazzo:2013uya,Bednyakov:2015sca}. The fate of the EW vacuum depends crucially on the Higgs mass (known now with a very good accuracy \cite{Tanabashi:2018oca}) and the top Yukawa coupling \cite{Krasnikov:1978pu,Politzer:1978ic,Hung:1979dn}: there is a critical value such that,
%if the top Yukawa is larger/smaller than it, the EW vacuum is meta-/absolutely stable.
if the top Yukawa is smaller (larger) than it, the EW vacuum is (meta)stable.
Reducing both theoretical and experimental uncertainties is the future challenge; see Ref. \cite{Bezrukov:2014ina} for a review and references therein.

If the SM itself can be extended up to very high energy scale, the question is, at which energy scale should we find deviations from the SM?\footnote{\label{footnote1}The SM cannot explain neutrino masses and oscillations, dark matter (see, however, Ref. \cite{Farrar:2017eqq}) and the baryon asymmetry of the Universe and thus cannot be a final theory.}
The answer to this question is unknown; several very different hypotheses, linked to the hierarchy problem,  have been put forward.

One possible approach to the stability of the EW scale against quadratically divergent radiative corrections which tend to bring the Higgs mass up to the scale of new physics, such as that of Grand Unified Theory (GUT), is associated with modifications of the SM right above the EW scale. This includes low-scale supersymmetry, extra dimensions or composite Higgs models (for reviews, see e.g. Refs. \cite{Chung:2003fi,Feng:2013pwa,Rubakov:2001kp,Davoudiasl:2009cd,Perelstein:2005ka,Panico:2015jxa}).
However, the fact that no convincing deviation from the SM has been observed so far places this conjecture under strain. 

In weakly coupled theories, it is the presence of very heavy particles like GUT leptoquarks with substantial interactions with the Higgs boson that jeopardizes the Higgs mass \cite{Gildener:1976ai}.
Thus, the requirement of the {\em nonexistence} of such particles may serve as a guiding principle for the construction of theories beyond the SM that address the hierarchy between the EW scale $\sim 100$ GeV and the Planck scale\footnote{Note that the Planck scale does not correspond to a mass of any elementary particle but merely serves as a measure of the strength of the gravitational interaction mediated by the massless gravitons.} $M_P  \equiv (8 \pi G_N )^{-1/2} \simeq 2.44 \times 10^{18}$GeV in a different manner \cite{Vissani:1997ys,Shaposhnikov:2007nj,Shaposhnikov:2009pv,Wetterich:2011aa,Farina:2013mla,Karananas:2017mxm}.
All the phenomenological drawbacks of the SM (see footnote \ref{footnote1}) can be cured by very feebly interacting particles with masses below the EW scale \cite{Asaka:2005an,Asaka:2005pn}  or, by heavier particles with somewhat larger coupling constants \cite{Farina:2013mla}.

Even if a theory does not contain any particles with the mass above the EW scale, it can enter into a strong coupling regime above a certain energy scale $\Lambda$.
This is the case for the SM coupled to gravity: depending on the magnitude of the nonminimal coupling $\xi$ of the Higgs field $h$ to the gravitational Ricci curvature scalar $R$, $\xi h^2 R$, the theory leaves the perturbative regime at a certain energy scale $E \sim M_P$ if $\xi \lesssim 1$, and at $E \sim M_P/\xi$ if $\xi \gg 1$, see e.g. Refs. \cite{Burgess:2009ea,Barbon:2009ya}.\footnote{These estimates are true for zero or small background values of the Higgs field $\ul{h}$.
When $\ul{h}$ is large, they are substantially modified; see Ref. \cite{Bezrukov:2010jz} and below.}
Entering into the strong coupling regime and violating the tree unitarity condition for the scattering amplitudes \cite{Cornwall:1974km} do not mean that the theory must be replaced by a new one with extra heavy fundamental degrees of freedom; it might ``self-complete'' or  ``self-heal'' itself  \cite{Weinberg:1980gg,Reuter:1996cp,Dvali:2010bf,Dvali:2010jz,Aydemir:2012nz}.
Whether the existence of such a strong coupling scale $\Lambda$ much higher than the EW scale leads to the hierarchy problem is uncertain due to the absence of reliable methods in the strong coupling regime.
A semiclassical approach to the hierarchy problem within Euclidean gravity \cite{Shaposhnikov:2018xkv,Shaposhnikov:2018jag} shows that the answer may be negative if the underlying theory is approximately Weyl invariant  for large values of the Higgs field.

It is interesting that a similar requirement of asymptotic ({\it large Higgs field}) scale invariance of the SM coupled to gravity  is crucial for Higgs inflation \cite{Bezrukov:2007ep,Bezrukov:2010jz}.
Also, classical scale invariance of the SM with vanishing Higgs mass has been invoked to address the hierarchy problem in numerous works (see e.g. Refs. \cite{Bardeen:1995kv,Meissner:2006zh,Foot:2007as,Hambye:2007vf,Iso:2009ss,Holthausen:2009uc,Hur:2011sv,Karam:2015jta,Wetterich:1983bi,Wetterich:2016uxm} and also, for models with dynamically induced Planck scale, e.g. Refs. \cite{Wetterich:1987fm,Foot:2007iy,Salvio:2014soa,Einhorn:2014gfa}).
Further, exact quantum scale invariance together with the absence of superheavy particles leads to the perturbative stability of the Higgs mass against radiative corrections \cite{Shaposhnikov:2008xi}. 

\

All these considerations motivated us to study the following questions.
Suppose that we modify the SM  in such a way that it becomes scale-invariant at the quantum level above a certain value of the Higgs field $h\sim \Lambda$.
This can be done, at least formally, with a specific scheme that we will present in this work or, equivalently, by adding an infinite number of higher-dimensional operators in a proper manner \cite{Mooij:2018hew}.
Then, the theory turns out to be strongly coupled at the energy scale $E \sim \Lambda$.
Can we still make perturbative computations of certain quantities? What could be the phenomenological and cosmological consequences of this theory?

The paper is organized as follows. In Sec. \ref{Asymptotic scale invariance}, we give a perturbative definition of asymptotically scale-invariant theories with the use of a simple toy model without gravity and discuss the associated strong coupling scale. 
Then in Sec. \ref{lambda stops ``running''}, the effective potential is computed based on the ordinary (justifiable) loop expansion.
In Sec. \ref{Asymptotic scale invariance with nonminimal coupling to gravity}, the toy model is extended with the nonminimal coupling to gravity and the strong coupling scale is reconsidered.
Then in Sec. \ref{lambda stops ``running'' before/after it jumps},
the effective potential is computed again with the ordinary method.
In Sec. \ref{Cosmological consequences},
considering the well-known properties of the SM,
we discuss cosmological implications  based on the effective potential obtained previously.
We summarize our prescription and findings in Sec. \ref{Summary}.
In the Appendix, we give some details of the perturbative computation of the effective potential.

%%%%%%%%%%%%%%%%%%%%%%%%%%%%%%%%%%%%%%%%%%%%%%%%
%%%%%%%%%%%%%%%%%%%%%%%%%%%%%%%%%%%%%%%%%%%%%%%%%%%%%%%%%%%%%%%%%%%%%%%%%%%%%%%%%%%%%%%%%%%%%%%%

\section{\label{Asymptotic scale invariance}Asymptotic scale invariance}

In this section, we give a perturbative definition of the class of quantum field theories preserving asymptotic scale invariance at the quantum level,
comparing it with the ``standard'' one breaking scale invariance and the one respecting exact scale invariance.
This class of theories involves a ``hidden'' mass parameter to be interpreted as a feature of the UV completion. We clarify how such an unconventional definition works in a self-consistent way.
We also discuss the energy scale below which perturbative computations are justified and the low-energy field theory description is reliable.

To clarify our point, we work with a simple Higgs-Yukawa model where a Dirac fermion $f$ interacts with a real scalar field $h$.
The classical action in four-dimensional flat spacetime is given by
\be
S_{\rm HY} = \int d^4 x~{\cal L}_{\rm HY}~, \label{S_HY}
\ee 
\be
-{\cal L}_{\rm HY} &=& \frac{1}{2} g^{\mu \nu} \partial_{\mu} h \partial_{\nu} h + V + \bar{f} \Slash{\partial} f + Y ~, \nn \\
Y &\equiv & y ~ h \bar{f} f /\sqrt{2} ~,~~~V \equiv \lambda~  h^{4}/4 ~,  \nn 
\ee
where $Y$ describes the Yukawa interaction through a coupling constant $y$ and $V$ is the quartic self-interaction term with a coupling constant $\lambda$.
Since there is no explicit mass scale in the Lagrangian, the model is (at least classically) scale-invariant:
the action (\ref{S_HY}) is invariant under the scaling transformation
\be
\begin{aligned}
x^{\mu} &\to  x'^{\mu} \equiv e^{-\sigma} ~ x^{\mu} ~,\\
\Phi(x) &\to  \Phi'(x') = e^{d_{\Phi} \sigma} ~ \Phi(x) ~,
\end{aligned}  \label{scaling-transformation-1}
\ee
where $\Phi$ stands for the each matter field and $d_{\Phi}$ denotes $\Phi$'s mass dimension: $d_h = 1$ and $d_f = 3/2$ in four-dimensional spacetime.

%%%%%%%%%%%%%%%%%%%%%%%%%%%%%%%%%%%%%%%
\subsection{\label{``Standard'' prescription}``Standard'' prescription}
In order to define a theory at the quantum level, we first need to regularize the model.
Throughout this paper, we employ dimensional regularization \cite{tHooft:1972tcz}; the model is extended to $n=(4-2\ve )$-dimensional spacetime where the mass dimensions of the dynamical fields $h$ and $f$ become
$d_h = (n-2)/2 = 1-\ve$ and $d_f = (n-1)/2 = 3/2 -\ve$, respectively.
Accordingly,
the mass dimensions of the quartic and the Yukawa couplings become $2 \ve$ and $\ve$, respectively.
These dimensionful ``bare'' couplings are written as
\be
\hat{\lambda} = \mu^{\frac{2\ve}{1-\ve}} \left[ \lambda + \sum_{i=0}^{\infty} \frac{{\sf C}^{\lambda}_{(i)}}{\ve^i} \right] 
, ~~~
\hat{y} =\mu^{\frac{\ve}{1-\ve}} \left[ y +  \sum_{i=0}^{\infty} \frac{{\sf C}^{y}_{(i)}}{\ve^i} \right] 
 \label{bare-couplings}
\ee
with an explicit scale $\mu$, the normalization point, whose mass dimension is assumed to be $1-\ve$ as is the case with the scalar field $h$, and the coefficients ${\sf C}^{\lambda /y}_{(i\geq 1)}$ properly chosen at each order of the perturbative computation to deal with the ultraviolet (UV) divergences.
And also, the dynamical fields in (\ref{S_HY}) need to be interpreted as ``bare'' fields for removing the divergences completely.
The finite parts ${\sf C}^{\lambda /y}_{(0)}$ are to redefine the couplings at each order of perturbative computation.

Here, as the standard prescription with the dimensional regularization, a ``minimal'' set of operators is assumed: we do not introduce new mass parameters aside from the normalization point $\mu$ in (\ref{bare-couplings}) and we respect the renormalizability of the model.
The only explicit mass parameter $\mu$ {\it breaks} scale invariance in $n$-dimensional spacetime, and then it is transmitted into the four-dimensional limit as the trace anomaly.

%%%%%%%%%%%%%%%%%%%%%%%%%%%%%%%%%%%%%%
\subsection{\label{Exactly scale-invariant prescription}Exactly scale-invariant prescription} 
 
Before going to the asymptotically scale-invariant prescription,
it is enlightening to look at the exactly scale-invariant (SI) case.
Instead of introducing an explicit mass scale $\mu$ to break scale invariance, one can identify a dynamical scalar field  as the normalization point \cite{Englert:1976ep} so that it also gets scaled under the transformation (\ref{scaling-transformation-1}).

Introducing a new dynamical scalar field $\phi$ besides the ``SM Higgs'' field $h$, replace $\mu$ with
\be
\omega \propto \phi \times F(x) ~, \label{omega_SI}
\ee
where $F$ is an arbitrary function of the dimensionless combination $x = h/\phi$ with ``hidden'' dimensionless parameters \cite{Shaposhnikov:2008xi} absent in the classical Lagrangian.
In general, any number of dynamical scalars can contribute to the normalization point.
And in any case, the vacuum state must break scale invariance spontaneously with finite expectation values of the scalars; otherwise, the perturbative expansion is ill defined.\footnote{If one takes the lattice regularization \cite{Shaposhnikov:2008ar} as a nonperturbative approach, the lattice spacing is proportional to $\omega^{-1}$.}

Because of the quantum fluctuations of the scalar fields in $\omega$, this class of theories is nonrenormalizable \cite{Shaposhnikov:2009nk}.
However, the renormalizability is not essential for the construction and scale invariance can be manifest at each order of the perturbative computation with counter terms respecting the symmetry.
One obtains the traceless energy-momentum tensor in flat spacetime but with the momentum dependent running couplings \cite{Armillis:2013wya,Gretsch:2013ooa,Tamarit:2013vda}.

Two prescriptions often used in the literature assume
\be
F(x) = \sqrt{1 + \Xi ~ x^2}  \label{F-omega_SI}
\ee
with (I) a particular value of $\Xi$ chosen in connection with a gravitational theory \cite{Shaposhnikov:2008xb,Shaposhnikov:2008xi,GarciaBellido:2011de,Bezrukov:2012hx,Rubio:2014wta,Trashorras:2016azl,Karananas:2016kyt,Shkerin:2016ssc,Tokareva:2017nng,Casas:2017wjh} (see footnote \ref{GR-SI} on page \pageref{GR-SI}) or (II)  $\Xi = 0$ so that the SM Higgs does not take part of the normalization point \cite{Ghilencea:2015mza,Ghilencea:2016ckm,Ghilencea:2016dsl,Ghilencea:2017yqv}.

%%%%%%%%%%%%%%%%%%%%%%%%%%%%%%%%%%%%%%%
\subsection{\label{Asymptotically scale-invariant prescription}Asymptotically scale-invariant prescription}

Let us introduce the asymptotically scale-invariant (aSI) 
prescription which preserves scale invariance only in the large field limit.
Here, we do not introduce the additional scalar.
However, it shares many features with the exactly SI prescription as seen below.

\

Working with dimensional regularization, we replace the constant $\mu$ in (\ref{bare-couplings}) by the field dependent normalization point (\ref{omega_SI}) but with a nondynamical mass scale instead of $\phi$: 
\be
\omega = \mu \times F(h/\mu_{\star}) ~, \label{omega_aSI}
\ee
where
$\mu_{\star}$ is a constant scale with the mass dimension $1-\ve$.
And here,
the function $F$ is to behave as
\be
F(x) \to \left\{ \begin{matrix} ~1  \\ ~x \end{matrix} \right. ~~{\rm for}~~ \begin{matrix} ~x\ll 1 \\ ~x\gg 1 \end{matrix} \nn 
\ee
so that $\omega$ is proportional to the dynamical scalar $h$ for
\be
h \gg \mu_{\star}  \label{large-field-1}
\ee
to recover the invariance under the scaling transformation (\ref{scaling-transformation-1}) in $n$-dimensional spacetime in the large field limit.
With (\ref{omega_aSI}), we have\footnote{Once a full theory has been specified,
one may scale the normalization point (\ref{omega_aSI}) and thus its overall factor $\mu$ with $\hat{\lambda}$, $\hat{y}$ and all the other bare quantities fixed.
We refer readers to Refs. \cite{Tamarit:2013vda,Lalak:2018bow} for discussions with exact scale invariance and also Refs. \cite{Ghilencea:2015mza,Ghilencea:2016ckm,Ghilencea:2016dsl,Ghilencea:2017yqv} for explicit computations of beta functions.}
\be
\begin{aligned}
\hat{\lambda}^{\rm aSI}  = \omega^{\frac{2\ve}{1-\ve}} \left[ \lambda + \sum_{i=0}^{\infty} \frac{{\sf C}^{\lambda}_{(i)}}{\ve^i} \right] = & ~F^{\frac{2\ve}{1-\ve}} \times \hat{\lambda} 
\ , \\
\hat{y}^{\rm aSI} = \omega^{\frac{\ve}{1-\ve}}\left[ y + \sum_{i=0}^{\infty} \frac{{\sf C}^{y}_{(i)}}{\ve^i} \right] =  & ~F^{\frac{\ve}{1-\ve}} \times \hat{y}
\end{aligned}
\label{aSI-bare-couplings}
\ee
instead of (\ref{bare-couplings}).
The $n$-dimensional interaction terms are now given as
\be
\hat{Y} \equiv  \omega^{\frac{\ve}{1-\ve}} Y ~~~,~~~ \hat{V}\equiv  ~  \omega^{\frac{2\ve}{1-\ve}} V  \label{n-dimensional-interaction-terms}
\ee
which, in the large field regime (\ref{large-field-1}), behave in the SI manner \cite{Englert:1976ep}:
\be
\hat{Y}|_{h \gg \mu_\star} \propto   h^{\frac{1}{1-\ve}} \overline{f} f ~~~,~~~ \hat{V}|_{h \gg \mu_\star} \propto  h^{\frac{4-2\ve}{1-\ve}} \label{SI-behavior}
\ee
whose mass dimension $n=4-2\ve$ is totally attributed to the dynamical fields.
As in the exact SI case, the quantum fluctuations of $\omega$ make the theory nonrenormalizable.
Therefore, the theory needs an infinite number of operators absent in (\ref{S_HY}) on which we impose respecting the symmetry.\footnote{Given a full Lagrangian,
one can make it look as if the normalization point is constant by expanding the field dependence in (\ref{aSI-bare-couplings}) with respect to $h/\mu_\star$ (see (\ref{standard-like}) in the Appendix).
In other words, it is possible to reproduce asymptotic scale invariance starting with the standard regularization with the explicit mass scale $\mu$ with higher-dimensional operators taken into account in a proper way.
One can find discussions in the literature, based on cutoff regularization \cite{Hamada:2016onh} or based on dimensional regularization \cite{Lalak:2018bow} for the exactly SI case.}

\

More explicitly, the aSI prescription we adopt here is as follows.
\begin{itemize}
\item The field dependent normalization point $\omega$ is asymptotically proportional to the dynamical field as (\ref{omega_aSI}).
We assume one of the simplest forms: 
\be
\omega = \mu \times \sqrt{ 1 + h^2 / \mu_{\star}^2 }  \label{omega}
\ee
with $\mu_{\star}$ being a free parameter.
\item All the ``higher-dimensional'' operators are suppressed by negative powers of
\be
\Lambda^{2}_{\star} \equiv \mu_{\star}^2 \times F^2(h/\mu_{\star}) = \mu_{\star}^2 + h^2  ~.  \label{Lambda}
\ee
Multiplied by appropriate powers of $\omega$ in $n$-dimensional spacetime,
such nonpolynomial operators behave as (\ref{SI-behavior}) in the large field regime (\ref{large-field-1}) to respect asymptotic scale invariance.
Their coefficients are set to cancel the divergences.
\end{itemize}
By expanding $F$ with respect to the quantum fluctuation of the scalar field $\delta h$ around a classical background $\ul{h}$ in a similar manner as (\ref{omega-expanded}) below,\footnote{In this paper, all the quantities evaluated with the homogeneous scalar field background are underlined.}
one can see that such nonpolynomial operators are enough.
The combination of these two manifests asymptotic scale invariance at each order of the perturbative computation.

For instance, the operators needed to cancel the divergences in the scalar potential\footnote{In general, scalar-fermion interactions and derivative operators suppressed by (\ref{Lambda}) are required to deal with divergent subdiagrams.
However, we do not need those for the two-loop level computation of the effective potential performed in this work.}
are written as
\be
\hat{W} =  \sum_{k=0}^{\infty} \frac{\hat{\lambda}_{[k]}^{\rm aSI}}{4} \frac{h^{4+2k}}{\Lambda_{\star}^{2k}} \label{W_HY}
\ee
with 
\be
\hat{\lambda}_{[k]}^{\rm aSI}  
=  F^{\frac{2\ve}{1-\ve}} \times \hat{\lambda}_{[k]} ~~ ,~~~ \hat{\lambda}_{[k]} = \mu^{\frac{2\ve}{1-\ve}} \left[ \lambda_{[k]} + \sum_{i=0}^{\infty} \frac{{\sf C}^{\lambda}_{[k](i)}}{\ve^i} \right]  ~, \nn
\ee
where the $k=0$ term corresponds to the ``quartic'' interaction. 
It should noted here that, for the SI behavior (\ref{SI-behavior}) to hold, this kind of summation of the infinite series needs to converge.
Such a nonperturbative issue is to be interpreted as a constraint on the UV completion (see also footnote \ref{convergence} on page \pageref{convergence}).

\

In addition,
we make the following assumption.
\begin{itemize}
\item
The tree level Lagrangian is simply given by 
\be
-{\cal L}_{{\rm HY}[0]} &=& \frac{1}{2} \eta^{\mu \nu} \partial_{\mu} h \partial_{\nu} h + \hat{V} +\overline{f} \Slash{\partial}  f  + \hat{Y}   \label{aSI-L_HY[0]}
\ee
with (\ref{n-dimensional-interaction-terms}).
We only add necessary nonpolynomial operators canceling the divergences at each order of the perturbative computation, whose finite parts
are sufficiently small as to be consistently omitted from lower-order computations.
\end{itemize}
With this assumption, we can work with a finite number of coupling constants \cite{Bezrukov:2014ipa,Ghilencea:2015mza,Ghilencea:2016ckm,Ghilencea:2016dsl,Ghilencea:2017yqv}.

%%%%%%%%%%%%%%%%%%%%%%%%%%%%%%%%%%%%%%%
\subsection{\label{Tree unitarity}Tree unitarity}
Now that the theory is nonrenormalizable,
there exists an energy scale above which the perturbative analyses fail.
As exploited in the context of Higgs inflation \cite{Bezrukov:2010jz},
so-called tree unitarity \cite{Cornwall:1974km} is a useful criterion for the validity, that is, $N$-particle tree amplitudes behave as
\be
{\cal M}_{N} \sim ({\rm energy~scale})^{q} ~~ {\rm with} ~~ q \leq 4-N ~. \nn
\ee
The $N$-scalar tree amplitude at $h=\ul{h}$ leads to the violation scale
\be
\Lambda_{N}  \sim \left[ \ul{\partial_{h}^N W} \right]^{\frac{-1}{N-4}} \nn 
\ee
for $N>4$
where $W$ is considered to have the same structure as the four-dimensional limit of (\ref{W_HY}).
Then, as naively expected from the suppression with (\ref{Lambda}),
we find the tree unitarity violation scale
\be
\Lambda_{\rm HY} \equiv {\rm min}_N \{ \Lambda_{N} \} \sim  \sqrt{\mu_\star^2 + \ul{h}^2} = \ul{\Lambda}_{\star} \label{tree-unitarity-violation-scale}
\ee
for a given background value $\ul{h}$, which has the same field dependence as the normalization point (\ref{omega}).
Nonrenormalizable operators other than $W$ also come with (\ref{Lambda}) to give rise to the same tree unitarity violation scale $\sim \ul{\Lambda}_{\star}$.
Note that we omitted the factors coming from the phase space integration and that is why the coupling constants did not appear in (\ref{tree-unitarity-violation-scale}).
With those factors, the couplings of the nonpolynomial operators that are small because of the assumption (\ref{aSI-L_HY[0]}) can make the tree unitarity violation scale higher by a factor of ${\cal O}(10)$ \cite{Dicus:2004rg}.
Hence, the expression (\ref{tree-unitarity-violation-scale}) is to be understood as a rather cautious bound.

\

We expect, therefore, that the complete Lagrangian of the aSI effective theory is given by (\ref{S_HY}) plus all sorts of ``higher-dimensional'' operators suppressed by negative powers of the field dependent scale (\ref{Lambda}).
These operators are capable to remove all UV divergencies in the theory and make the approach self-consistent (see Refs. \cite{Bezrukov:2010jz,Bezrukov:2014ipa} for discussions in the context of Higgs inflation).
In our ``bottom-up'' approach, it is impossible to make any definite statement about physics above the tree unitarity violation scale (see, however, Refs. \cite{Weinberg:1980gg,Reuter:1996cp,Dvali:2010bf,Dvali:2010jz,Aydemir:2012nz} mentioned in Sec. \ref{Introduction}).
Nevertheless,
if all the coefficients in front of these operators were known, the theory would be predictive for all energy scales, at least in principle.
Whether the theory has an admissible behavior in the  UV domain is an open question.
Within the spirit of asymptotic safety \cite{Weinberg:1980gg,Reuter:1996cp}, we will assume that this is indeed the case at least with some choice of the dimensionless coefficients.

In spite of the presence of the tree unitarity violation scale,
certain quantities, such as the effective potential, can still be computed reliably. Indeed, here the relevant energy scales are masses of particles in the vacuum diagrams. In our  model (\ref{S_HY}), the fermion and the scalar masses are $m_f = y \ul{h}/\sqrt{2}$ and $m_h = \sqrt{3 \lambda } \ul{h}$, respectively. For small couplings  we always have $m_{f/h} < \Lambda_{\rm HY}$, and, hence, the perturbative computation of the effective potential is possible. The point that the UV cutoff is dependent on the scalar field background is crucial for our discussions below.\footnote{For example, adding to the action all sorts of higher-dimensional operators suppressed by the tree unitarity violation scale {\it evaluated with vanishing scalar background}, $\ul{\Lambda}_{\star}\vert_{\ul{h}=0} = \mu_{*}$, as has been advocated in Refs. \cite{Burgess:2009ea,Barbon:2009ya}, would make the effective potential noncomputable for large values of the Higgs field.}

%%%%%%%%%%%%%%%%%%%%%%%%%%%%%%%%%%%%%%%
\section{\label{lambda stops ``running''}Application : $\bm \lambda$ stops ``running''}
Let us compute the effective potential of the classical homogeneous background $\ul{h}$ perturbatively with the aSI prescription and the assumption (\ref{aSI-L_HY[0]}) to check its SI asymptotic behavior $\propto \ul{h}^4$.

\

All differences stem from having the dynamical quantity $\omega$ instead of $\mu$.
It becomes obvious how this replacement affects the effective potential once $\omega$ is expanded with respect to the quantum fluctuation of the scalar field $\delta h$ around $\ul{h}$ as
\be
\omega^{\frac{2 q \ve}{1-\ve}}&=&\ul{\omega}^{\frac{2 q \ve}{1-\ve}} ~{\Bigg [}1+ \sum_{i\geq j}^{\infty} {\cal E}^{(q)}_{i,j} ~ \frac{\ul{h}^{j}  \delta h^{i}}{\ul{\Lambda}_{\star}^{i+j}} {\Bigg ]} ~, \label{omega-expanded}
\ee
where $q=1,1/2$ for the scalar potential and the Yukawa, respectively, and the terms with the coefficients ${\cal E}^{(q)}_{k,j}$ (for the definition, see (\ref{Epsilon-odd}) and (\ref{Epsilon-even}) in the Appendix) are ``evanescent'' or ``abnormal''
to vanish in the four-dimensional limit $\ve \to 0$.
First of all, we should find $\ul{\omega}$ where the constant $\mu$ usually is with the ``standard'' prescription: it appears as $[\ln (m_{f/h}^2 / \ul{\omega}^2)]^{i\geq 1}$.
Second, the evanescent contributions multiplied by the $1/\ve$ divergence can contribute as finite nonpolynomial corrections in the four-dimensional limit.
Beyond the one-loop level,
multiplied by $1/\ve^{i\geq 2}$, the evanescent terms can give rise to the divergences with $\ul{\Lambda}_{\star}^{- 2k}$ to require the $k\geq 1$ terms in (\ref{W_HY}); see Refs. \cite{Ghilencea:2015mza,Ghilencea:2016ckm,Ghilencea:2016dsl,Ghilencea:2017yqv} for computations with the exactly SI prescription.

To be more explicit,
the effective potential at the one-loop level is given as
\be
{\bm V}_{\rm aSI} = \ul{V} + {\cal V}_1(\ul{h};\ul{\omega})+  U_{1\star}(\ul{h}) ~,  \nn
\ee
where 
\be
{\cal V}_1 (\ul{h};\ul{\omega}) \equiv \frac{\ul{h}^{4}/4}{(4 \pi)^2} \sum_{x=h,f} \frac{B^{x}}{2} \ln \left(\frac{m_{x}^2}{\ul{\omega}^2 e^{3/2}} \right)  \label{Coleman-Weinberg-1-aSI}
\ee
is the Coleman-Weinberg correction \cite{Coleman:1973jx} now with $\ul{\omega}$ and the coefficients are $B^{h} = 18 \lambda^2$ and $B^{f} = -2y^4$; additionally,
\be
U_{1\star}(\ul{h}) = -\frac{m_h^4 m_{h,1}^{2}}{2 (4 \pi)^2} \equiv  \sum_{k=1}^{2} ~ \frac{A_{[k]}}{(4\pi)^{2}} ~ \frac{\ul{h}^{4+2k}}{4 \ul{\Lambda}_{\star}^{2k}} \label{U_1star}
\ee
with $A_{[1]} = -27 \lambda^2$ and $A_{[2]} = 6 \lambda^2$
comes from the evanescent contribution to the scalar mass squared:
\be
\ul{\omega}^{\frac{-2 \ve}{1-\ve}} \ul{\partial_h^2 \hat{V}} &=& m_h^2 + \ve \times m_{h,1}^2 + {\cal O}(\ve^2) ~ , \nn \\
 m_{h,1}^2 &\equiv & m_h^2 \times \left[ \frac{3 \ul{h}^2}{2\ul{\Lambda}_{\star}^2} - \frac{\ul{h}^4 }{3 \ul{\Lambda}_{\star}^4} \right] ~. \nn
\ee
The fermion mass, on the other hand, has no evanescent term.
Furthermore, the two-loop level computation (following e.g. Ref. \cite{Ford:1992pn}) explicitly shows that the correction $\sim \ul{h}^4 [\ln (m_{x}^2 / \ul{\omega}^2)]^{2}$ appears and that the nonpolynomial contributions like (\ref{U_1star}) also come with $\ln (m_{x}^2 / \ul{\omega}^2)$; see Appendix \ref{Two-loop level corrections}.

Higher powers of $\ln (m_x^2/\ul{\omega}^2)$ as well as higher powers of the ``suppression'' factor $\ul{h}^2 /\ul{\Lambda}_{\star}^{2} = (1 + \mu_{\star}^2/\ul{h}^2)^{-1}$ are expected in higher-loop corrections.
Considering all those terms,
we define a field dependent quantity ${\bm \lambda}_{\rm aSI}$ as
\be
{\bm \lambda}_{\rm aSI}(\ul{h};\mu_\star;\mu) &\equiv & \frac{{\bm V}_{\rm aSI}}{\ul{h}^4 /4} \nn \\
&\equiv & \sum_{k=0}^{\infty} \frac{\ul{h}^{2k}}{ \ul{\Lambda}_{\star}^{2k} } ~{\bm \lambda}_{[k]}(\ul{h};\mu_\star;\mu) \label{effective-lambda_aSI}  \\
&=& \sum_{k,l=0}^{\infty} \frac{\ul{h}^{2k}}{ \ul{\Lambda}_{\star}^{2k} } \frac{{\mf B}_{[k]}^{(l)}}{l! (4 \pi)^{2l}} \left(\ln \frac{\ul{h}}{\ul{\omega}}\right)^l  \nn  \\ 
&=&  \sum_{k,l=0}^{\infty}\frac{{\mf B}_{[k],\infty}^{(l)}}{l! (4 \pi)^{2l}} \frac{\left\{ - \frac{1}{2} \ln (1 + \mu_\star^2 /\ul{h}^2 ) \right\}^l}{\left( 1 + \mu_\star^2 /\ul{h}^2 \right)^k} ~, \nn
\ee
where ${\mf B}^{(l)}_{[k]} \equiv (4 \pi)^{2l}  \left[ \{-\mu \partial_{\mu} \}^{l} {\bm \lambda}_{[k]}(\ul{h};\mu_\star;\mu) \right]_{\ul{h}=\ul{\omega}}$ on the third line and 
${\mf B}^{(l)}_{[k] , \infty}  \equiv  (4 \pi)^{2l}   \{-\mu \partial_{\mu} \}^{l} {\bm \lambda}_{[k]}(\infty ;\mu_\star;\mu)$ on the fourth line.\footnote{\label{infinity-mu_star}Formally, ${\bm \lambda}_{[k]}(\infty ;\mu_\star;\mu) = {\bm \lambda}_{[k]}(\mu_\star ; \infty;\mu)$ holds since $\ul{h}/\ul{\omega}$ approaches $\mu_\star /\mu $ in the large field limit.}
Now,
the quantity (\ref{effective-lambda_aSI}) loses its field dependence for $\ul{h} \gg \mu_\star$ to asymptotically approach
\be
{\bm \lambda}_{\rm SI}(\mu_\star ; \mu) \equiv {\bm \lambda}_{\rm aSI}(\infty ;\mu_\star;\mu) = \sum_{k=0}^{\infty}{\mf B}_{[k],\infty}^{(0)}  ~, \label{lambda_SI}
\ee
which is to say,
the effective scalar potential in the large field regime (\ref{large-field-1}) behaves in the scale-invariant way $\propto \ul{h}^4 $ as if there were no quantum correction:\footnote{\label{convergence} At a finite order of the perturbative expansion, due to the assumption (\ref{aSI-L_HY[0]}), only a finite number of the coefficients ${\mf B}_{[k\geq 0],\infty}^{(0)}$ contribute to the summation in the asymptotic value (\ref{lambda_SI}).
We assume that the unknown UV completion guarantees the convergence of the summation and thus the finiteness of (\ref{lambda_SI}), leading to the aSI low-energy effective theory.}
\be
{\bm V}_{\rm aSI} \approx {\bm \lambda}_{\rm SI}(\mu_\star ; \mu) \times \ul{h}^4 ~.  \label{SI-regime-potential}
\ee

\

\begin{figure}[t]
\centering
\includegraphics[width=1 \linewidth, bb=0 0 553 309]{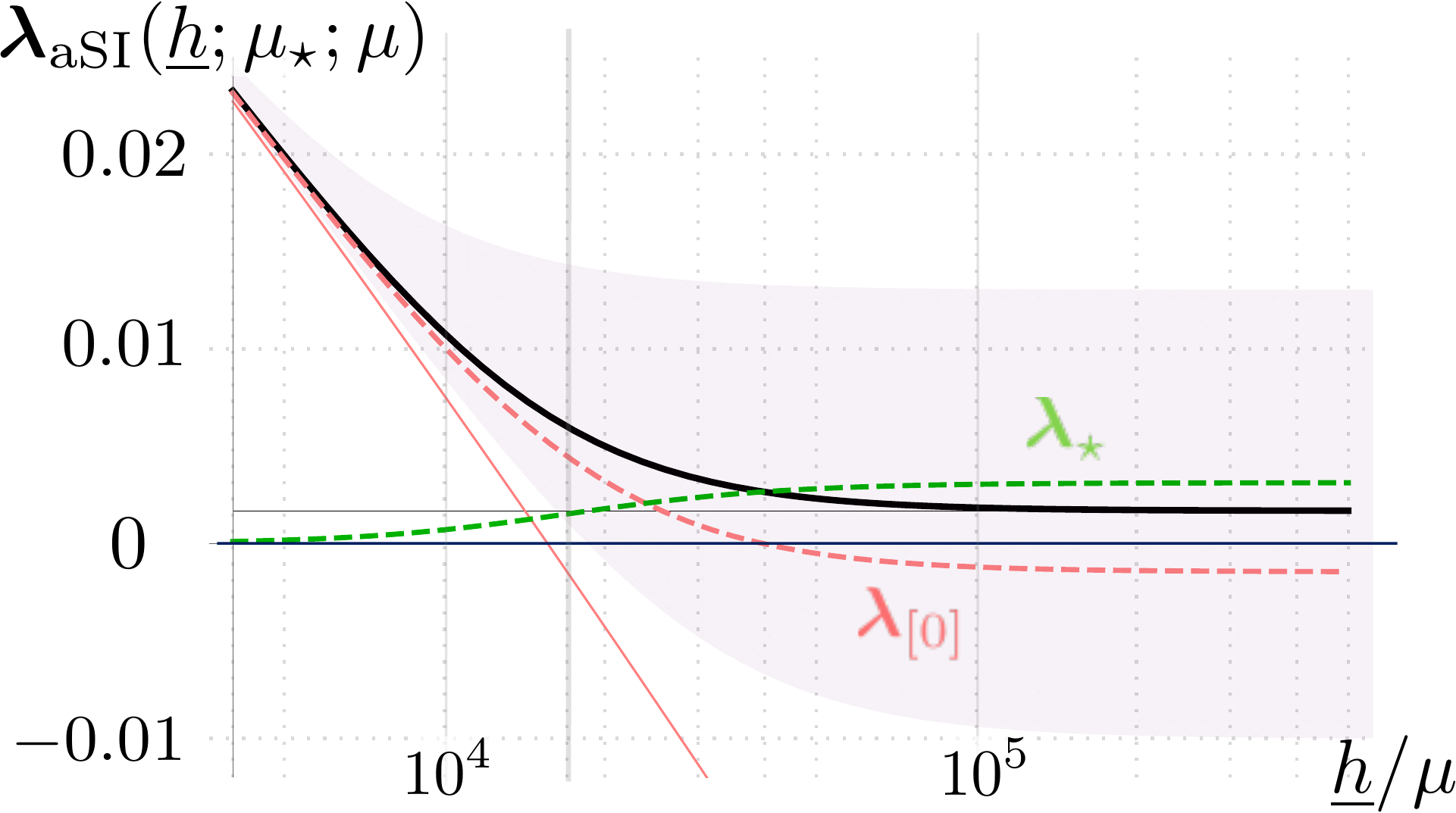}
\caption{
\label{fig1}The field dependent quantity (\ref{effective-lambda_aSI}) at the two-loop level is plotted with the black line for $\lambda =0.12$, $y =1$ and $\mu_\star /\mu =1.7 \times 10^4$.
The asymptotic value (\ref{lambda_SI}) is shown by the thin horizontal line.
The red dashed and green dashed lines show ${\bm \lambda}_{[0]}$ and ${\bm \lambda}_\star$, respectively.
The red thin line is the ``standard'' result (\ref{effective-lambda_standard}) for comparison.
The shaded region is swept downward (upward) by the black line of ${\bm \lambda}_{\rm aSI}$ when $\mu_\star$ is continuously changed to be doubled (halved).
Here, the finite parts ${\sf C}^{\lambda}_{[k](0)}$ are fixed in the $\overline{\rm MS}$-scheme manner for definiteness (see Appendix \ref{Two-loop level corrections} for some details).
}
\end{figure}

In order to make a comparison with the ``standard'' result,
let us take the limit of $\mu_\star \to \infty$ where $\omega$ coincides with $\mu$.
The counterpart of (\ref{effective-lambda_aSI}) is then obtained as
\be
{\bm \lambda}(\ul{h};\mu) &\equiv &{\bm \lambda}_{\rm aSI}(\ul{h};\infty ;\mu)  = {\bm \lambda}_{[0]}(\ul{h};\infty;\mu) \nn \\
&=& \sum_{l=0}^{\infty} \frac{{\mf B}_{[0]}^{(l)}}{l! (4 \pi)^{2l}} \left(\ln \frac{\ul{h}}{\mu}\right)^l ~ . \label{effective-lambda_standard}
\ee
If $2 m_f^2 > m_h^2$, the ``beta function'' ${\mf B}_{[0]}^{(1)} \approx B^h + B^f$ is negative,
and then, ${\bm \lambda}$ decreases and crosses zero at the scalar field value $h_\star$ satisfying 
\be
{\bm \lambda}(h_{\star};\mu) = 0 ~~~~{\rm with} ~~~ \begin{aligned} &- \partial_{\ul{h}}{\bm \lambda}(\ul{h};\mu)|_{\ul{h} = h_{\star}} \\
&~~~= \partial_{\mu}{\bm \lambda}(h_{\star};\mu)   > 0 \end{aligned}  ~~. \label{h_star}
\ee
However, now with the aSI prescription,
$\ul{h} / \ul{\omega}$ asymptotically approaches $\mu_\star / \mu$ and 
${\bm \lambda}_{[0]}$ goes to
\be
{\bm \lambda}_{[0]}(\infty ;\mu_\star;\mu) = {\bm \lambda}(\mu_\star;\mu) \nn 
\ee
which does not cross zero if $\mu_\star < h_\star$.
In addition, the rest of (\ref{effective-lambda_aSI}),
\be
{\bm \lambda}_{\star}(\ul{h};\mu_\star;\mu) \equiv {\bm \lambda}_{\rm aSI}(\ul{h};\mu_\star;\mu) - {\bm \lambda}_{[0]}(\ul{h};\mu_\star;\mu) ~, \label{lambda_star}
\ee
has a finite asymptotic value which can be positive while loop suppressed with our assumption (\ref{aSI-L_HY[0]}).
Figure \ref{fig1} shows a two-loop level example where $\mu_\star$ is such close to $h_\star$ that the asymptotic values of ${\bm \lambda}_{[0]}$ and ${\bm \lambda}_\star$ are comparable and the asymptotic value (\ref{lambda_SI}) becomes positive if $\mu_\star$ is sufficiently small.
%%%%%
%%%%%

\

Let us emphasize that the above is a reliable result: it has been derived in the weakly coupled regime in that the masses of particles in the vacuum diagrams are below the tree unitarity violation scale, as noted at the end of Sec. \ref{Tree unitarity}.

Since the structure (\ref{effective-lambda_aSI}) does not depend on field contents, 
applied to the SM, asymptotic scale invariance drastically changes the shape of the Higgs effective potential.
Especially,
even with the current {\it central} experimental value of the top quark mass,
{\it absolute} stability of the EW vacuum is realized with $\mu_\star \sim 10^{10}$GeV being of the order of the energy scale at which the Higgs quartic coupling would cross zero in the canonical SM.

%%%%%%%%%%%%%%%%%%%%%%%%%%%%%%%%%%%%%%%
%%%%%%%%%%%%%%%%%%%%%%%%%%%%%%%%%%%%%%%
%%%%%%%%%%%%%%%%%%%%%%%%%%%%%%%%%%%%%%%

\section{\label{Asymptotic scale invariance with nonminimal coupling to gravity}Asymptotic scale invariance\\ with nonminimal coupling to gravity}

In this section, we discuss asymptotic scale invariance with the nonminimal coupling to gravity.
After reviewing well-known facts associated with the frame change,
we apply the aSI prescription introduced in Sec. \ref{Asymptotically scale-invariant prescription} in the Jordan frame and then move to the Einstein frame.
Also, we discuss the energy scale below which our low-energy description is valid.

\

We add a gravitational part to the Higgs-Yukawa action (\ref{S_HY})  as
\be
S= \int d^4 x \sqrt{-g} ~ \left( {\cal L}_{R} + {\cal L}_{\rm HY} \right) ~, \label{S}
\ee
\be
{\cal L}_{R} = \frac{M_{P}^2 + \xi h^2}{2} R ~, \nn 
\ee
where $R$ is the Ricci scalar with respect to the spacetime metric $g_{\mu\nu}$ and $M_P$ is the only explicit mass scale at the classical level.
${\cal L}_{\rm HY}$ is as in (\ref{S_HY}) but with the partial derivatives replaced by the covariant ones.
The scalar field $h$ has a positive nonminimal coupling $\xi$ to gravity.

%%%%%%%%%%%%%%%%%%%%%%%%%%%%%%%%%%%%%%%
\subsection{\label{Mass scales and symmetry of classical theory}Mass scales and symmetry of classical theory}
The combination
\be
M_{P,{\rm eff}}^2 \equiv M_P^2 + \xi h^2 \label{effective-Planck}
\ee
is to be interpreted as the effective Planck mass squared
which is field dependent.
For this field dependence, $g_{\mu\nu}$ is called the Jordan frame metric.
Due to the nonminimal coupling,
$M_P$ becomes negligible when the scalar field value $h$ is much larger than
\be
\mu_{\rm si} \equiv  \frac{M_P}{\sqrt{\xi}} \ . \label{mu_si}
\ee
Therefore, the model is aSI, at least at the classical level:
in the large scalar field regime
\be
h \gg \mu_{\rm si} ~, \label{large-field-2}
\ee
the action (\ref{S}) is approximately invariant under the scaling transformation (\ref{scaling-transformation-1}) with the metric unchanged.
Thanks to diffeomorphism invariance, it is equivalent to the symmetry under the change of variables as
\be
\begin{aligned}
g_{\mu \nu}(x) &\to e^{-2 \sigma} ~ g_{\mu \nu}(x) ~ ,\\
\Phi(x) &\to e^{\hat{d}_{\Phi} \sigma} ~ \Phi(x) ~,
\end{aligned}  \label{scaling-transformation-2}
\ee
where $\hat{d}_{\Phi}$ coincides with the mass dimension in the Higgs-Yukawa model: $\hat{d}_{\Phi} = d_{\Phi}$.

\ 

One can switch to the Einstein frame to work with the field independent Planck scale.
The spacetime metric in this frame is, in four-dimensional spacetime, related to the one in the Jordan frame by the Weyl transformation
\be
\wt{g}_{\mu\nu} = \Omega^2 ~ g_{\mu\nu} ~ ,\nn
\ee
\be
\Omega^2 \equiv 1 + \frac{\xi h^2}{M_P^2} = 1 + \frac{h^2}{\mu_{\rm si}^2}  ~. \nn 
\ee
Accordingly,
the field dependence of the Planck mass (\ref{effective-Planck}) disappears as $M_{P,{\rm eff}}/\Omega = M_P$.
Also, the Weyl-transformed fermion field $\wt{f} = f/\Omega^{3/2}$ turns out to be canonically normalized.
On the other hand, the canonically normalized scalar field, denoted by $\chi$, is a nontrivial mixture of the scalar field $h$ and the metric in the Jordan frame through the nonminimal coupling.
As discussed in Sec. \ref{From Jordan to Einstein},
when $h$ is larger than
\be
\mu_{\rm km} \equiv  \frac{M_P}{\sqrt{6 \xi^2 + \xi}} = \frac{\mu_{\rm si}}{\sqrt{6 \xi + 1}} \leq \mu_{\rm si} ~, \label{mu_km}
\ee
$\chi$ differs significantly from $h$, not to mix with the Einstein frame metric.
And especially, for (\ref{large-field-2}),
it is approximately written as \cite{Bezrukov:2007ep,Bezrukov:2008ut} 
\be
\chi 
\approx \sqrt{6} ~ M_P \times \ln (h / \mu_{\rm si}) ~. \label{chi-h-relation}
\ee

Based on these canonically normalized fields, the invariance under the scaling transformation (\ref{scaling-transformation-2}) in the Jordan frame is equivalent to shift symmetry with
\be
\chi(x) \to \chi(x) + {\rm const} ~, \label{shift}
\ee
but with the other fields unchanged:
\be
\wt{g}_{\mu \nu}(x) \to \wt{g}_{\mu \nu}(x) ~~ , ~~~ \wt{f}(x) \to \wt{f}(x) ~. \nn
\ee
The scalar field $\chi$ is nothing but the pseudo Nambu-Goldstone boson associated with asymptotic scale invariance: the dilaton.

%%%%%%%%%%%%%%%%%%%%%%%%%%%%%%%%%%%%%%%
\subsection{\label{From Jordan to Einstein}Asymptotically scale-invariant prescription: \\ From Jordan to Einstein}
There exist two conventional prescriptions in the literature.
The first option \cite{Bezrukov:2007ep} is known as prescription I which preserves asymptotic scale invariance.
And the second one \cite{Barvinsky:2008ia} is prescription II which does not preserve asymptotic scale invariance.
As seen below and summarized in Table \ref{table-prescriptions},
each option corresponds to a special value of $\mu_\star$ even though it is {\it a~priori} an arbitrary mass scale.
Here we generalize the conventional prescriptions with $\mu_\star$ as a free parameter.

\

Let us start by defining the theory in the Jordan frame in the same way as in Sec. \ref{Asymptotically scale-invariant prescription} with the field dependent normalization point $\omega = \mu \sqrt{1 + h^2 /\mu_\star^2}$. 
Unlike the matter fields, the mass dimensions of the metric and the Ricci scalar do not depend on the spacetime dimension.
With the explicit mass scale $M_P$ extended to have the same mass dimension as the scalar field $1-\ve$ in $n=(4-2\ve)$-dimensional spacetime, the nonminimal coupling is always dimensionless.
Therefore, only the couplings in ${\cal L}_{\rm HY}$ are replaced with the dimensionful ones in (\ref{aSI-bare-couplings}).
So here, as the tree level Lagrangian in the Jordan frame,
we have 
\be
{\cal L}_{[0]} =\frac{\Omega^2   M_P^2 }{2} \, R + {\cal L}_{{\rm HY}[0]} ~, \label{Jordan-aSI-L_[0]}
\ee
where ${\cal L}_{{\rm HY}[0]}$ is nothing but the general covariant version of (\ref{aSI-L_HY[0]}).
Then, approximate scale invariance holds, not for (\ref{large-field-2}), but for
\be
h \gg {\rm max}\{ \mu_{\rm si}, \mu_\star \} ~. \label{large-field}
\ee

\

Let us switch to the Einstein frame.
The $n$-dimensional metric is defined as
\be
\wt{g}_{\mu\nu} =  \Omega^{\frac{4}{n-2}} ~ g_{\mu \nu} = \Omega^{\frac{2}{(1-\varepsilon)}} ~ g_{\mu \nu}  \nn
\ee
from which the Ricci scalar $\wt{R}$ and the covariant derivative $\wt{\nabla}$ in the Einstein frame are constructed.
The Einstein frame Lagrangian is written as
\be
\wt{\cal L}_{[0]} =\frac{M_P^2 }{2} \, \wt{R} + \wt{\cal L}_{{\rm HY}[0]}  ~, \nn
\ee
where
\be
-\wt{\cal L}_{{\rm HY}[0]} &=& {\cal G}_h  ~ \frac{1}{2} \wt{g}^{\mu \nu} \partial_{\mu} h \partial_{\nu} h + \hat{\wt{V}} +\wt{\bar{f}} \Slash{\wt{\nabla}}  \wt{f}  + \hat{\wt{Y}} ~, \label{Einsein-aSI-L_HY[0]}
\ee
\be
\hat{\wt{Y}} \equiv \wt{\omega}^{\frac{\ve}{1-\ve}} \wt{Y}~~ , ~~~
\hat{\wt{V}}\equiv \wt{\omega}^{\frac{2\ve}{1-\ve}} \wt{V}~. \nn 
\ee
While the canonically normalized fermion is simply defined as
$\wt{f} = f/ \Omega^{d_f \frac{2}{n-2}}$,
the noncanonical factor for the scalar field
\be
{\cal G}_h &=& \frac{1}{\Omega^2}\left[ 1 + \frac{6 \xi }{\vs ~ \Omega^2} \frac{h^2}{\mu_{\rm si}^2} \right] \label{field-space-metric} \\
&=& \frac{1 + h^2 / \mu_{\rm km}^2}{\Omega^4}\left[ 1 + \frac{\ve ~ \xi /3}{\xi + 1/6} \frac{h^2}{\mu_{\rm km}^2 + h^2} + {\cal O}(\ve^2) \right] \nn 
\ee
with\footnote{
$\vs \ne 1$ simply reflects the fact that the conformal value of the nonminimal coupling in $n$-dimensional spacetime $\xi_c = -(1/4) (n-2)/(n-1)=-\vs /6$ is not $-1/6$.
While the abnormal term in (\ref{field-space-metric}) is absent if one extends the theory to $n$-dimensional spacetime after switching to the Einstein frame,
then a corresponding abnormal term in the Jordan frame is required for consistency.}
$\vs = (1-\ve)/(1-2\ve /3 )$ 
defines the canonically normalized scalar field $\chi$ by
\be
\frac{\partial \chi}{\partial h} = \sqrt{{\cal G}_h}  \nn 
\ee
which tells us that, when the scalar field is larger than $\mu_{\rm km}$ given as (\ref{mu_km}),
the mixing between the scalar field and the metric in the Jordan frame is significant to make the difference between $h$ and $\chi$.
As for the interaction terms,
not only the parts
\be
\wt{Y} \equiv \frac{Y}{\Omega^4} = \frac{y}{\sqrt{2}} ~ \frac{h}{\Omega} ~ \wt{\bar{f}}\wt{f} ~~,~~~ \wt{V} \equiv \frac{V}{\Omega^4} = \frac{\lambda}{4} ~  \frac{h^{4}}{\Omega^4} \label{Einstein-interaction-terms}
\ee
lose their $\chi$ dependences in the large field regime (\ref{large-field-2}) where $h/\Omega \to \mu_{\rm si}$,
but also the normalization point in the Einstein frame
\be
\wt{\omega} \equiv \frac{\omega}{\Omega} = \mu \times \sqrt{\frac{ 1 + h^2/\mu_\star^2 }{1 + h^2/\mu_{\rm si}^2} } \label{Einstein-omega}
\ee
approaches a constant $\mu$ for (\ref{large-field}) to be $\chi$ independent.
Therefore, as expected from the aSI nature of the tree level Lagrangian (\ref{Jordan-aSI-L_[0]}) in the Jordan frame,
asymptotic shift symmetry under (\ref{shift}) in the Einstein frame holds in $n$-dimensional spacetime.
In Table \ref{table-prescriptions}, two choices of $\mu_\star$ corresponding to prescriptions I and II are summarized.\footnote{\label{GR-SI}In the exactly SI case \cite{Shaposhnikov:2008xb,Shaposhnikov:2008xi,GarciaBellido:2011de,Bezrukov:2012hx,Rubio:2014wta,Trashorras:2016azl,Karananas:2016kyt,Shkerin:2016ssc,Tokareva:2017nng,Casas:2017wjh}, the Planck mass $M_P$ in (\ref{effective-Planck}) is replaced by an additional scalar field $\phi$ with a nonminimal coupling: $M_P^2 \to \xi' \phi^2$. 
The counterpart of prescription I is called the GR-SI prescription where the normalization point $\omega$ is proportional to the effective Planck mass $\sqrt{\xi' \phi^2 + \xi h^2}$ in the Jordan frame. This corresponds to the special choice $\Xi = \xi/\xi'$ in (\ref{F-omega_SI}).
See also Refs. \cite{Ferreira:2016vsc,Ferreira:2016wem,Ferreira:2016kxi,Ferreira:2018itt,Ferreira:2018qss} for discussions based on the conserved Weyl current.}
\begin{table}[t]
\centering 
%\begin{ruledtabular}
\begin{tabular}{cccc}
\hline \hline
Prescription & $\mu_\star$ & $\omega$ (Jordan) & ~~$\wt{\omega}$ (Einstein)~~ \\
\hline
I& $\mu_{\rm si}$& $\mu \times \Omega \propto M_{P,{\rm eff}}$ & $\mu$ : {\it constant} \\
II & $\infty$ & $\mu$ : {\it constant} & $\mu / \Omega$ \\ \hline \hline
\end{tabular}
\caption{\label{table-prescriptions}%
The two conventional values of $\mu_\star$ and the corresponding normalization points $\omega$ and $\wt{\omega}$.
}
%\end{ruledtabular}
\end{table}

Now that asymptotic shift symmetry is respected by the tree level Lagrangian to start with,
the UV divergences also respect the symmetry and can be removed without violating it; see \cite{Bezrukov:2010jz} for a discussion with prescription I.
As discussed in Sec. \ref{Asymptotically scale-invariant prescription},
we take the prescription that manifests the symmetry of the full quantum theory at each order of the perturbative computation and introduce only necessary operators for canceling the divergences.
As is clear from how the five explicit mass scales, $\mu_{\star/{\rm km}/{\rm si}}$, $M_P$ and $\mu$, appear in the tree level Lagrangian, only the three $\mu_{\star/{\rm km}/{\rm si}}$ are to be compared with $h$.
This means that the full theory is assumed to have approximate scale invariance (shift symmetry) in the Jordan (Einstein) frame in the large field regime (\ref{large-field}) as long as $\mu_\star$ is finite.
In the following, we do not assume the special choices of $\mu_\star$ in Table \ref{table-prescriptions}.

\

It should be noted that the quantum fluctuations of the Einstein frame metric require ``higher-dimensional'' terms in the scalar potential with $[(h^2/\Omega^2)/M_{P}^2]^{k\geq 1}$
$= [h^2/M_{P,{\rm eff}}^2]^{k\geq 1}$.
Those terms are strongly suppressed for $h\ll M_{P}$ and asymptotically behave as $\hat{\wt{V}}/\xi^{k \geq 1}$.
Therefore, neglecting the metric fluctuations in the Einstein frame can be justified even in the large field regime if $\xi > 1$.

%%%%%%%%%%%%%%%%%%%%%%%%%%%%%%%%%%%%%%
\subsection{\label{Tree-unitarity-with-xi}Tree unitarity}
First, for small values of the nonminimal coupling, we have $M_P < \mu_{\rm km} \approx \mu_{\rm si}$.
Then, for $h < \mu_{\rm si}$, the model is essentially the same as the model discussed in Sec. \ref{Asymptotic scale invariance} but now with gravity.
With $\Lambda_P$ being the tree unitarity violation scale associated with the metric fluctuations, we identify
\be
\wt{\Lambda}_{\rm HY}^{\xi < 1} = {\rm min}\{ \ul{\Lambda}_{\star}, \Lambda_P \} \label{tree-unitarity-violation-scale-for-xi<1}
\ee
as the tree unitarity violation scale.
For definiteness, let us apply the value of $\Lambda_P$ obtained in Ref. \cite{Han:2004wt} which is about one-half of $G_N^{-1/2} \simeq 1.2 \times 10^{19}$GeV for the SM particle contents.\footnote{The self-healing mechanism for the graviton-mediated scattering process was also identified in Ref. \cite{Han:2004wt}.}
We argue that (\ref{tree-unitarity-violation-scale-for-xi<1}) applies also to the large field regime $\ul{h} > \mu_{\rm si}$ because, if $\mu_\star > \Lambda_P$, no mass scale exists below (\ref{tree-unitarity-violation-scale-for-xi<1}); if $\mu_\star < \Lambda_P$, it should be already irrelevant to dynamics in the large field regime $\ul{h} > \mu_{\rm si} > \mu_{\star}$.

\

\begin{figure}[t]
\centering
\includegraphics[width=1 \linewidth, bb=0 0 641 403]{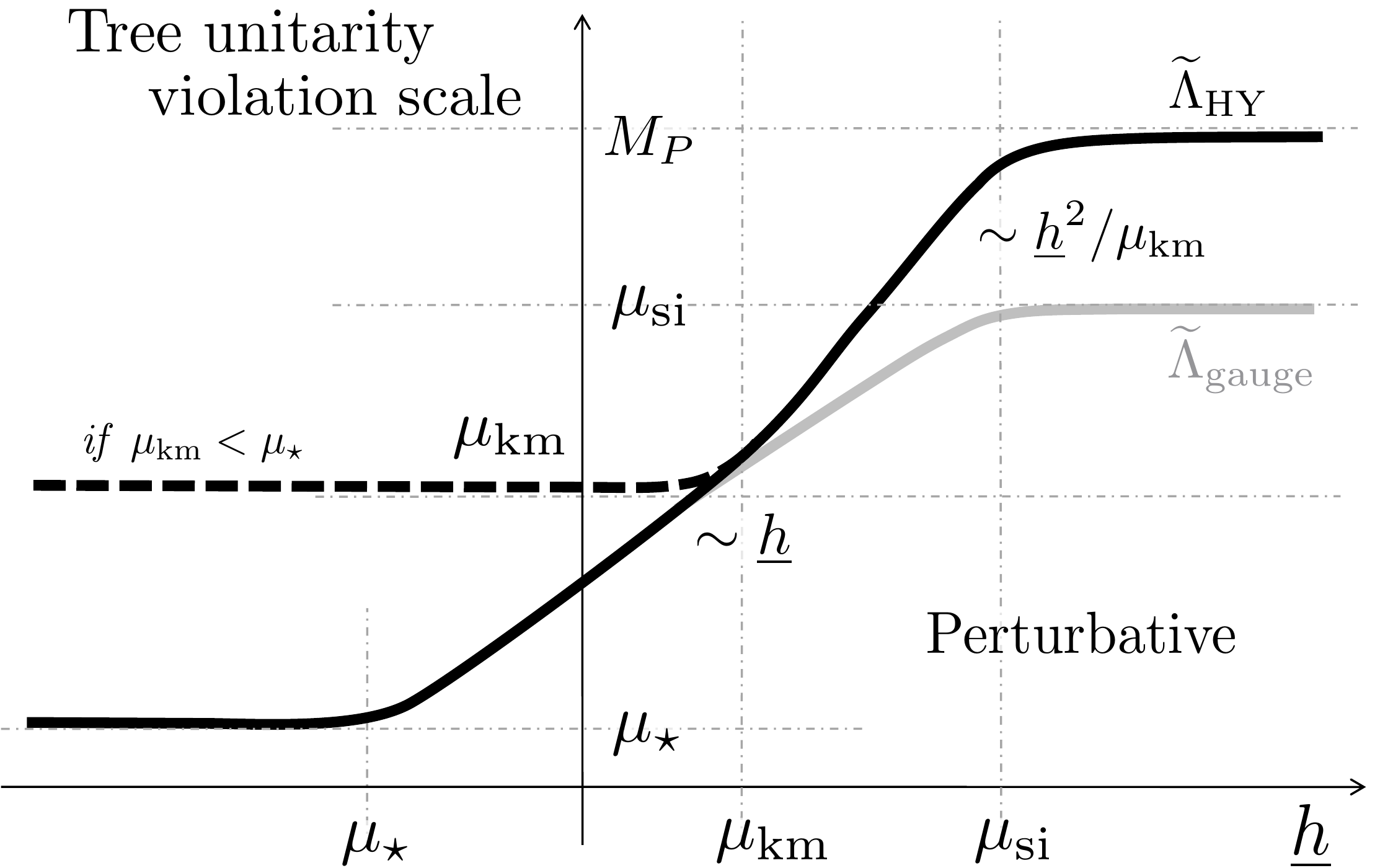}
\caption{\label{Fig:tree-unitarity} The tree unitarity violation scale with $\xi > 1$ in the Einstein frame. The solid (dashed) black line shows the field dependence of (\ref{wt-Lambda_HY}) with $\mu_\star$ smaller (larger) than $\mu_{\rm km}$. When $h$ is identified as the SM Higgs, the gauge interaction gives rise to (\ref{field-dependent-cutoff}) to be the lowest scale for $\ul{h} \gtrsim \mu_{\rm km}$ as depicted with the gray line.}
\end{figure}

For large values of the nonminimal coupling,
in order to take into account the Higgs-graviton mixing in the Jordan frame for $\ul{h} \gtrsim \mu_{\rm km}$ correctly,
we repeat the analysis in Sec. \ref{Tree unitarity} but with the three mass scales $\mu_{\star / {\rm km} /{\rm si}}$ and the noncanonical factor ${\cal G}_h$ and then compare with $M_P$ due to the metric fluctuations.
Let us identify the lowest energy scale of
\be
\wt{\Lambda}_{N}  \sim \left[ \ul{\partial_{\chi}^N \wt{W}} \right]^{\frac{-1}{N-4}} \label{wt-E^N}
\ee
for $N>4$, where 
\be
\wt{W} &=& \sum_{k,l,m} \frac{\lambda_{[k,l,m]} h^4}{4 \Omega^4} \left(\frac{h^2}{\mu_\star^2 +h^2}\right)^{k} \left(\frac{h^2}{\mu_{\rm km}^2+h^2}\right)^{l} \left(\frac{h^2}{\mu_{\rm si}^2+h^2}\right)^{m} \nn
\ee
is a general form of the asymptotically shift symmetric scalar self-interaction term in the four-dimensional spacetime with the three mass scales to be compared with $h$.
The $N$th derivative with respect to the canonically normalized scalar $\chi$ is rewritten as
\be
\partial_{\chi}^N = \left[ {\cal G}_h^{-1/2} \partial_h \right]^N 
= {\cal G}_h^{-\frac{N}{2}} \left[ \partial_h - (N-1) \frac{\partial_h {\cal G}_h}{2 {\cal G}_h} \right] \cdots  \left[ \partial_h -\frac{\partial_h {\cal G}_h}{2 {\cal G}_h} \right] \partial_h ~, \nn
\ee
where ${\cal G}_h$ is interpreted as the four-dimensional limit of (\ref{field-space-metric}) and
\be
\frac{\partial_h {\cal G}_h}{2 {\cal G}_h} = \frac{h}{\mu_{\rm km}^2 + h^2} - \frac{2h}{\mu_{\rm si}^2 + h^2}  \nn
\ee
is nothing but the field-space connection compatible with the metric ${\cal G}_h$.
Note that, acting on $\wt{W}$ repeatedly, the derivative $\partial_h$ and this connection both give the same structure.
Therefore, (\ref{wt-E^N}) is bounded below as
\be
\wt{\Lambda}_{N}  \gtrsim  \ul{{\cal G}_h}^{\frac{N/2}{N-4}}  \left[ \ul{\partial_{h}^N \wt{W}} \right]^{\frac{-1}{N-4}} ~.\label{wt-E^N-2} 
\ee
Also note that $\wt{W}$ has the same structure as $W$ which gives the violation scale (\ref{tree-unitarity-violation-scale}) in Sec. \ref{Tree unitarity} but now with the three different mass scales,
and then the lowest scale
\be
\mu_{-} \equiv {\rm min}\{ \mu_\star , \mu_{\rm km} , \mu_{\rm si}  \} ={\rm min}\{ \mu_\star , \mu_{\rm km} \} \label{mu_-}
\ee
replaces $\mu_\star$ in (\ref{tree-unitarity-violation-scale}).
For $\ul{h} \lesssim M_P$,
we have ${\cal G}_h>1$ to find
\be
\wt{\Lambda}_{\rm HY}^{\xi > 1} \equiv \ul{{\cal G}_h}^{1/2} \sqrt{\mu_-^2 + \ul{h}^2 } = \wt{\Lambda}_-   \sqrt{\frac{1+\ul{h}^2 /\mu_{\rm km}^2}{1+\ul{h}^2 /\mu_{\rm si}^2}}  \label{wt-Lambda_HY}
\ee
as a lower bound of (\ref{wt-E^N}) which reaches the constant $\mu_{\rm si}^2/\mu_{\rm km} \sim M_P$ already at $\ul{h} \sim \mu_{\rm si}$.
Here,
\be
\wt{\Lambda}_- \equiv \frac{\sqrt{\mu_-^2 + \ul{h}^2 }}{\ul{\Omega}}   \label{field-dependent-cutoff}
\ee
has been introduced.
Note that (\ref{wt-Lambda_HY}) turns out to be true even for $\ul{h} \gtrsim  M_P $ because one can find a lower bound stronger than (\ref{wt-E^N-2}) by following the relationship (\ref{chi-h-relation}) between $h$ and $\chi$;
each $\chi$ derivative gives rise to $1/M_P$, which means $\wt{\Lambda}_N \gtrsim M_P$ and the expression (\ref{wt-Lambda_HY}) is applicable.
This is lower than or the same as $M_P$ and, hence identified as the tree unitarity violation scale for $\xi > 1$.
In Fig. \ref{Fig:tree-unitarity}, the field dependence of (\ref{wt-Lambda_HY}) is depicted with the black lines.
If $\mu_{\rm km} < \mu_\star $,
it coincides with the one derived in \cite{Bezrukov:2010jz}.
The Yukawa interaction does not change the result. 
Multiplied by $\ul{\Omega}$, it gives the tree unitarity violation scale in the Jordan frame.

\

When we discuss the effective potential, the tree unitarity violation scale is to be compared with masses of particles in the vacuum diagrams.
As seen shortly later,
the Einstein frame masses are smaller than $\ul{h}/\ul{\Omega}$.
If $\xi > 1$, the tree unitarity violation scale (\ref{wt-Lambda_HY}) and even (\ref{field-dependent-cutoff}) turn out to be higher than the masses irrespective of the scalar field value:
the perturbative computation of the effective potential is justified even for trans-Planckian field values.
On the other hand, if $\xi < 1$, the masses $\propto \ul{h}/\ul{\Omega}$ eventually exceed $\Lambda_P \simeq 6 \times 10^{18}$GeV and thus (\ref{tree-unitarity-violation-scale-for-xi<1}) at a certain scalar field value $h_{\rm tuv}$ which can be larger than $\Lambda_P$ when the couplings are small.
Then, the naive perturbative computation becomes unreliable for $\ul{h} > h_{\rm tuv} > \Lambda_P$.

\

In Sec. \ref{Cosmological consequences}, we identify the scalar field $h$ as the SM Higgs boson which has gauge interactions.
As discussed in Ref. \cite{Bezrukov:2010jz},
the interaction between the canonically normalized Higgs excitation and the gauge boson becomes weaker, and, hence,
the Higgs mechanism breaks down to violate tree unitarity around the energy scale $\ul{h}/\ul{\Omega}$ corresponding to the mass of the gauge boson divided by the gauge coupling.
And also, the nonpolynomial interaction with $\mu_\star$ leads to this breakdown.
Then, we have 
\be
\wt{\Lambda}_{\rm gauge}  \sim  \wt{\Lambda}_- ~, \nn 
\ee
which is lower than (\ref{wt-Lambda_HY}) for $\xi > 1$ as seen in Fig. \ref{Fig:tree-unitarity} depicted with the gray line.
For $\mu_{\star} < \mu_{\rm km}$, we find it proportional to the normalization point (\ref{Einstein-omega}).
As the expression applicable also to the case with $\xi < 1$,
we regard ${\rm min}\{\Lambda_{-}, \Lambda_P \}$ as the tree unitarity violation scale.

%%%%%%%%%%%%%%%%%%%%%%%%%%%%%%%%%%%%%%%
\section{\label{lambda stops ``running'' before/after it jumps}Application :\\ $\bm \lambda$ stops ``running'' before/after it jumps}
Let us compute the effective potential perturbatively in the Einstein frame with the aSI prescription assuming flat spacetime.
The one-loop computation is enough to see peculiar behaviors including the threshold correction discussed in Ref. \cite{Bezrukov:2014ipa,Bezrukov:2017dyv},
while we make brief comments on higher-loop corrections.
We do not specify the value of the nonminimal coupling here,
having in mind that, if $\xi < 1$, the validity holds only when particles' masses are smaller than $\Lambda_P$; more explicitly, only when $\ul{h} < h_{\rm tuv}$, where $h_{\rm tuv}^2 = \Lambda_P^2 / {\rm max}\{ y^2/2 , 3 \lambda \}$ in the Higgs-Yukawa model.

\
 
The tree level mass of the canonically normalized fermion $\wt{f}$ is simply given as
\be
\ul{\wt{\omega}}^{\frac{-\ve}{1-\ve}} \partial_{\wt{f}} \partial_{\wt{\bar{f}}} \ul{\hat{\wt{Y}}} = \frac{ m_{f}}{ \ul{\Omega} } ~, \nn
\ee
where $m_f = y \ul{h}/\sqrt{2}$ is the Jordan frame mass.
On the other hand, 
the tree level mass squared of the canonically normalized scalar fluctuation around $\ul{\chi}$ comes with the evanescent contribution as
\be
\ul{\wt{\omega}}^{\frac{-2 \ve}{1-\ve}} \ul{ \partial_{\chi}^2 \hat{\wt{V}}} =  \frac{m_{h}^2}{\ul{\Omega}^2} {\Bigg [} z +  \frac{\ve ~\vt_{\vs} z_{\vs}}{1 + \mu_{\rm km}^2/\ul{h}^2} +  \frac{\ve ~ \vt_\star  z_\star}{1 + \mu_\star^2/\ul{h}^2} + {\cal O}(\ve^2) {\Bigg ]} ~, \nn
\ee
where $m_h^2 = 3 \lambda \ul{h}^2$ and
\be
z \equiv \frac{1}{ 1+ \ul{h}^2 /\mu_{\rm km}^2 } \left( a - \frac{ b~ \ul{h}^2}{\mu_{\rm km}^2 + \ul{h}^2} - \frac{c~ \ul{h}^2}{\mu_{\rm si}^2 + \ul{h}^2} - \frac{d~ \ul{h}^2}{\mu_{\star}^2 + \ul{h}^2} \right) \nn 
\ee
with $(a,b,c,d)=(1,1/3,4/3,0)$ comes from the noncanonical factor (\ref{field-space-metric}) and the denominator $\Omega^4$ in (\ref{Einstein-interaction-terms}) and suppresses $\chi$'s mass squared by a factor $\sim \mu_{\rm km}^2 /\ul{h}^2$ for $\ul{h} \gtrsim \mu_{\rm km}$.
The abnormal contribution of ${\cal O}(\ve)$ is divided into two parts:
the one with
\be
\vt_{\vs} \equiv \frac{\xi}{\xi +1/6} = 1-\frac{\mu_{\rm km}^2}{\mu_{\rm si}^2} \nn
\ee
comes from $\vs-1 = -\ve /3 + {\cal O}(\ve^2)$ in (\ref{field-space-metric})
to vanish in the minimally coupled limit $\xi \to 0$, whereas the other with
\be
\vt_\star \equiv 1-\frac{\mu_\star^2}{\mu_{\rm si}^2} \nn
\ee
is due to the field dependence of (\ref{Einstein-omega}) which is absent if one takes prescription I.
Both come with suppression factors $z_{\vs}$ and $z_{\star}$ similar to $z$ but with $(a,b,c,d)=(-4/9,-2/9,-4/9,0)$ and $(a,b,c,d)=(3/2,1/6,4/3,1/3)$, respectively.
Besides, they also have the extra suppression factor $(1 + \mu_{{\rm km}/\star}^2/\ul{h}^2)^{-1}$ for the small field regime $\ul{h} \lesssim \mu_{{\rm km}/\star}$.

\

Unlike the case discussed in Sec. \ref{lambda stops ``running''},
the one-loop vacuum bubble of the scalar field is nonrenormalizable already at the one-loop level because of the factor $z$.
Introducing a new operator necessary for eliminating the divergence,
we get the Einstein frame effective potential
\be
\wt{\bm V}_{\rm aSI} = \frac{\ul{V}+{\cal V}_{1}^{f}(\ul{h};\ul{\omega}) + {\cal V}_{1}^{\chi}(\ul{h};\ul{\omega}) + u_{\vs} (\ul{h}) + u_{\star} (\ul{h})}{\ul{\Omega}^4}  \nn 
\ee
at the one-loop level, where
\be
{\cal V}_1^{f} (\ul{h};\ul{\omega})  &\equiv & \frac{\ul{h}^4}{4 (4 \pi)^2} \left\{ -{\mf C} + \frac{B^{f}}{2} \ln \left( \frac{m_f^2}{\ul{\omega}^2 e^{3/2}} \right) \right\}~,~~\label{V_1^f} \\
{\cal V}_1^{\chi} (\ul{h};\ul{\omega}) & \equiv & \frac{ \ul{h}^4 z^2}{4 (4 \pi)^2} \left\{ {\mf C} + \frac{B^{h}}{2} \ln \left( \frac{m_h^2 z}{\ul{\omega}^2 e^{3/2}} \right) \right\} ~~, \label{V_1^chi} \\
u_{\vs/\star} (\ul{h}) & \equiv & -\frac{B^{h}  \ul{h}^4}{4 ~ (4\pi)^2}\frac{\vt_{\vs/\star} z_{\vs/\star}  z}{1+\mu_{{\rm km}/\star}^2/\ul{h}^2}  \label{u_vs/star} 
\ee
with $B^f$ and $B^h$ given in (\ref{Coleman-Weinberg-1-aSI}).
$\mf C$ in (\ref{V_1^chi}) is introduced as a finite part of the operator associated with the divergence of $\chi$'s bubble; accordingly, $-\mf C$ in (\ref{V_1^f}) is also introduced for $\wt{\bm V}_{\rm aSI}$ to have the same expression as ${\bm V}_{\rm aSI}$ in the small field limit $\ul{h}\ll \mu_{\rm km}$ where $z \approx 1$ and $\ul{\Omega}\approx 1$: (\ref{V_1^f})+(\ref{V_1^chi}) corresponds to (\ref{Coleman-Weinberg-1-aSI}), the abnormal correction $u_{\star}$ reduces to (\ref{U_1star}) and $u_{\vs}$ disappears.\footnote{Similar considerations can apply for the kinetic terms and the Yukawa interaction term. However, at the one-loop level, they do not affect the effective potential.}
Note also that, for the arguments of the logarithms, one finds ratios of the tree level squared masses in the Einstein frame to $\ul{\wt{\omega}}^2= \ul{\omega}^2/\ul{\Omega}^2$ in the first place; then, cancellations of the $\ul{\Omega}$ factors lead to (\ref{V_1^f}) and (\ref{V_1^chi}).

\

Let us define a field dependent quantity ${\bm \lambda}_{\rm aSI}^{\xi}$ by dividing the effective potential by $\ul{h}^4/4$:
\be
{\bm \lambda}_{\rm aSI}^{\xi}(\ul{h};\mu_\star,\mu_{\rm km},\mu_{\rm si} ; \mu) 
={\bm \lambda}_{\rm ch} + {\bm \lambda}_{\rm km} + {\bm \lambda}_{\vs}  + {\bm \lambda}_{\wt{\star}} ~,~~~ \label{effective-lambda_aSI^xi}
\ee
where 
\be
&&{\bm \lambda}_{\rm ch}(\ul{h};\mu_\star; \mu) \equiv  \left[ \ul{V} + {\cal V}_{1}^{f}(\ul{h};\ul{\omega})\right]/(\ul{h}^4/4) ~, \nn \\
&&{\bm \lambda}_{\rm km}(\ul{h};\mu_\star,\mu_{\rm km},\mu_{\rm si} ; \mu) \equiv  {\cal V}_{1}^{\chi}(\ul{h};\ul{\omega})/(\ul{h}^4/4)~, \nn \\
&&{\bm \lambda}_{\vs /\wt{\star}}(\ul{h};\mu_\star,\mu_{\rm km},\mu_{\rm si} ; \mu)  \equiv  u_{\vs /\star}(\ul{h}) /(\ul{h}^4/4) \nn
\ee
at the one-loop level.

First of all,
the field dependence of ${\bm \lambda}_{\rm ch}$ disappears for $\ul{h} \gg \mu_\star$ in the same manner as in Sec. \ref{lambda stops ``running''} without the nonminimal coupling.
This respects the asymptotic symmetry under the shift (\ref{shift}) in the Einstein frame and, equivalently, asymptotic scale invariance in the Jordan frame.
The mass scale $\mu_{\rm si}$ from the effective Planck mass (\ref{effective-Planck}) is irrelevant here because of the cancellation of the $\ul{\Omega}$ factors mentioned above.

Second, in (\ref{V_1^chi}),
the extra field dependence encoded in $\ln z$ itself becomes significant for $\ul{h} \gtrsim \mu_{\rm km}$.
However, the prefactor $z^2$ has a stronger field dependence $\sim \mu_{\rm km}^4/\ul{h}^4$ to make ${\bm \lambda}_{\rm km}$ negligible compared to ${\bm \lambda}_{\rm ch}$ and consistent with asymptotic scale invariance.\footnote{If one reads off the beta function of the ``quartic'' coupling from the combination ${\bm \lambda}_{\rm ch} + {\bm \lambda}_{\rm km}$, it changes suddenly at $\ul{h} \sim \mu_{\rm km}$; see e.g. Refs. \cite{George:2013iia,George:2015nza}.}

And third, ${\bm \lambda}_{\vs}$ and ${\bm \lambda}_{\wt{\star}}$  are also suppressed by $\mu_{\rm km}^4/\ul{h}^4$ for $\ul{h} \gtrsim \mu_{\rm km}$.
In addition, each has the suppression factor $(1 + \mu_{{\rm km}/\star}^2/\ul{h}^2)^{-1}$ for small field values.
Thus, if ${\bm \lambda}_{\vs}$ impacts on the quantity (\ref{effective-lambda_aSI^xi}), it occurs only at $\ul{h} \sim \mu_{\rm km}$.
As for ${\bm \lambda}_{\wt{\star}}$, it can be significant for $\mu_{\star} \lesssim \ul{h} \lesssim \mu_{\rm km}$ only when $\mu_{\star} < \mu_{\rm km}$ holds.

\

\begin{figure}[t]
\centering
\includegraphics[width=1 \linewidth, bb=0 0 548 323]{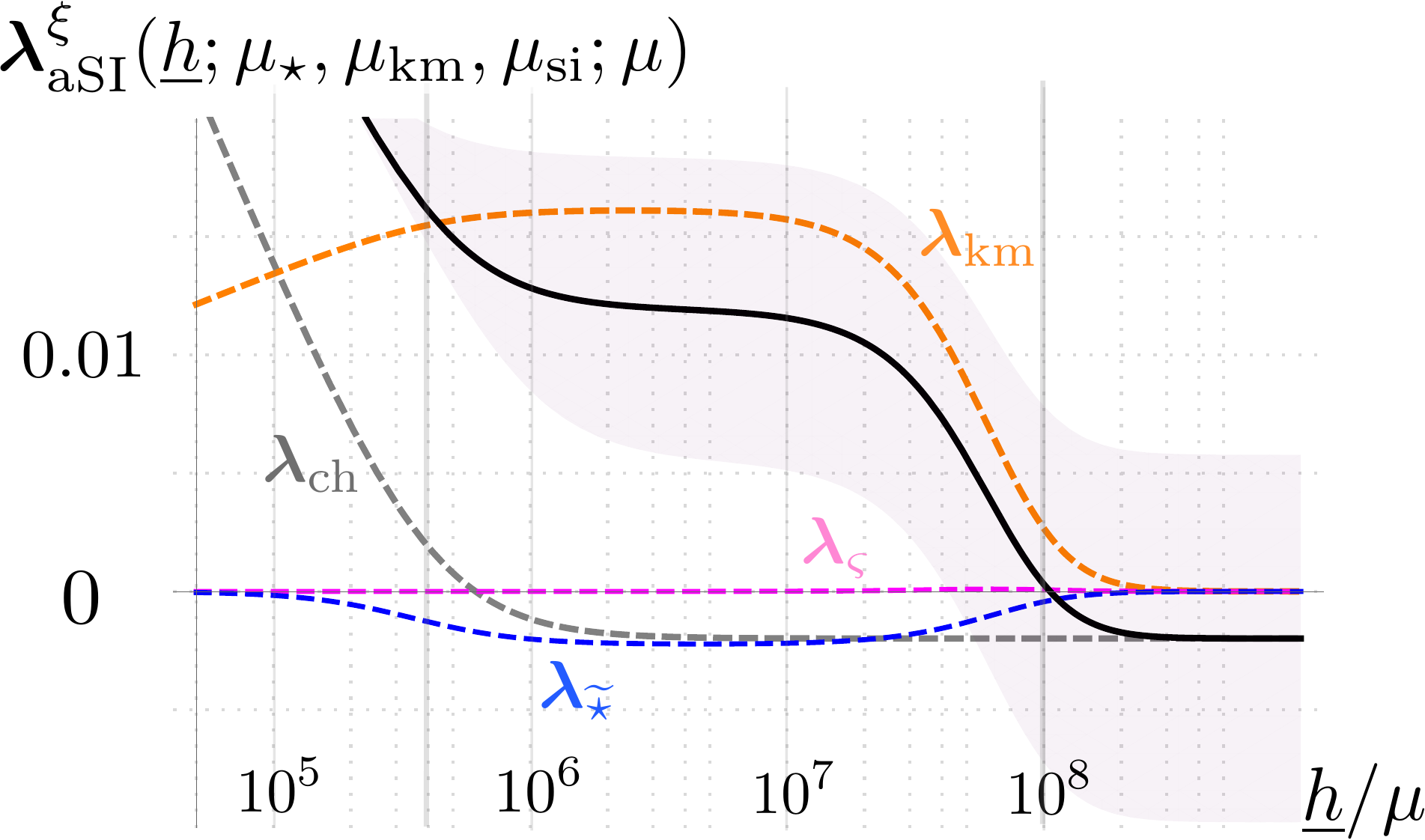}
\caption{\label{Fig:lambda_aSI^xi-1} The field dependent quantity ${\bm \lambda}_{\rm aSI}^{\xi}$ defined by (\ref{effective-lambda_aSI^xi}) is plotted with the black line for the dimensionless parameters $\lambda =0.13$, $y =0.97$ and ${\mf C} = -1$.
The dashed lines show the four contributions to it.
While their ``runnings'' disappear at $\ul{h} \sim \mu_{\star} = 4 \times 10^5 \, \mu$, the asymptotic value of ${\bm \lambda}_{\rm aSI}^{\xi}$ is reached for $\ul{h} > \mu_{\rm km} \approx 10^8 \, \mu$.
For our parameter choice, (\ref{Jump}) is positive even though ${\bm \lambda}_{\rm km}+ {\bm \lambda}_{\star}$ is negative at $\ul{h}\sim \mu$ with ${\mf C} < 0$.
The asymptotic value is given by (\ref{lambda_SI^xi}), which depends neither on $\mu_{\rm km}$ nor on $\mu_{\rm si} = 10^{10}\, \mu$.
The shaded region is swept downward (upward) by the black line of ${\bm \lambda}_{\rm aSI}^{\xi}$ when $\mu_\star$ is continuously changed to be doubled (halved).}
\end{figure}

\begin{figure}[t]
\centering
\includegraphics[width=1 \linewidth, bb=0 0 552 323]{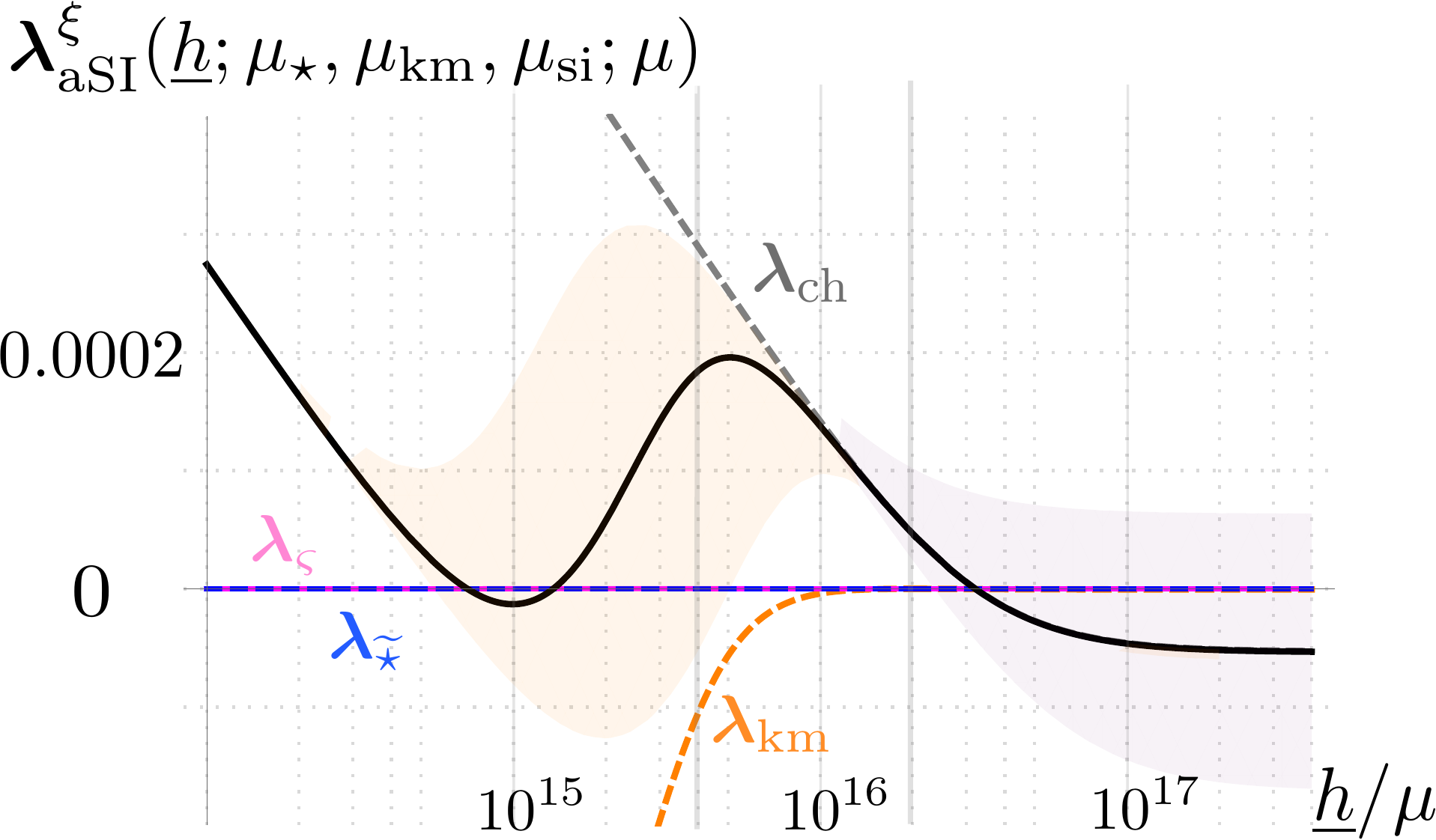}
\caption{\label{Fig:lambda_aSI^xi-2}The same quantities as in Fig. \ref{Fig:lambda_aSI^xi-1} but with different values of the parameters at $\mu$: $\lambda  =0.006$, $y=0.34$, ${\mf C} = -0.12$, $\mu_\star /\mu =2 \times 10^{16}$, $\mu_{\rm km}/\mu =4 \times 10^{15}$ and $\mu_{\rm si}/\mu = 2 \times 10^{16}$.
Here (\ref{Jump}) is negative, and, hence, ${\bm \lambda}_{\rm aSI}^{\xi}$ jumps up at $\ul{h}\sim \mu_{\rm km}$.
The gray-shaded region is swept downward (upward) by the black line of $\lambda^{\xi}_{\rm aSI}$ when $\mu_{\star}$ is continuously changed to be doubled (halved), and so is the orange-shaded region with $\mu_{\rm km}$ instead.}
\end{figure}

Two typical behaviors of ${\bm \lambda}_{\rm aSI}^{\xi}$ with $\mf C < 0$ are depicted in Figs. \ref{Fig:lambda_aSI^xi-1} and \ref{Fig:lambda_aSI^xi-2}.
In Fig. \ref{Fig:lambda_aSI^xi-1},
$\mu_\star \ll \mu_{\rm km}$ is assumed.
For $\ul{h} \ll \mu_{\rm km}$, the model is equivalent to the one without the nonminimal coupling.
Hence, ${\bm \lambda}_{\rm aSI}^{\xi}$ is approximated by ${\bm \lambda}_{\rm aSI}$ obtained in Sec. \ref{lambda stops ``running''} and becomes almost constant (\ref{lambda_SI}).
However, for $\ul{h} \gtrsim \mu_{\rm km}$, only the contribution of ${\bm \lambda}_{\rm ch}$ survives.
Therefore, we have
\be
{\bm \lambda}_{\rm aSI}^{\xi} \sim \left\{ \begin{matrix} {\bm \lambda}_{[0]}(\ul{h};\mu_\star ; \mu)  &  & \ul{h} \lesssim \mu_{\star}  \\
{\bm \lambda}_{\rm aSI}(\infty;\mu_\star ; \mu)  & \quad {\rm for} \quad & \mu_{\star} \lesssim \ul{h} \lesssim \mu_{\rm km}\\
{\bm \lambda}_{\rm ch}(\infty;\mu_\star ; \mu)  &  & \mu_{\rm km} \lesssim \ul{h} 
\end{matrix} \right. \label{rough-behavior-1}
\ee
with $\mu_\star \ll \mu_{\rm km}$.
Here, ${\bm \lambda}_{\rm aSI}= {\bm \lambda}_{[0]} + {\bm \lambda}_{\star}  \approx {\bm \lambda}_{[0]}$ for $\ul{h} \ll \mu_\star$ is applied.
${\bm \lambda}_{\rm aSI}^{\xi}$ jumps down (up) at $\ul{h} \sim \mu_{\rm km}$ when 
\be
-\Delta(\mu_\star,\mu_{\rm km},\mu_{\rm si} ; \mu)\equiv \left[ {\bm \lambda}_{\rm km} + {\bm \lambda}_{\star}\right]_{\ul{h}\sim \mu_{\rm km}} \label{Jump}
\ee
is positive (negative).
Meanwhile, ${\bm \lambda}_{\vs}$ may contribute there to make a wiggle 
of the size of $\delta \equiv |{\bm \lambda}_{\vs}|_{\ul{h}\sim \mu_{\rm km}}$.
%%%%
%%%%%
On the other hand, in Fig. \ref{Fig:lambda_aSI^xi-2}, we assume the opposite, $\mu_{\rm km}\ll \mu_\star$. The quantity ${\bm \lambda}_{\rm aSI}^{\xi}$ is still ``running'' at $\ul{h}\sim \mu_{\rm km}$ and, hence,  
\be
{\bm \lambda}_{\rm aSI}^{\xi} \sim \left\{ \begin{matrix} {\bm \lambda}_{[0]}(\ul{h};\mu_\star ; \mu)  &  & \ul{h} \lesssim \mu_{\rm km}  \\
{\bm \lambda}_{\rm ch}(\ul{h};\mu_\star ; \mu)  & \quad {\rm for} \quad & \mu_{\rm km} \lesssim \ul{h} \lesssim \mu_{\star}\\
{\bm \lambda}_{\rm ch}(\infty;\mu_\star ; \mu)  &  & \mu_{\star} \lesssim \ul{h} 
\end{matrix} \right. ~. \label{rough-behavior-2}
\ee
The conventional prescriptions in Table \ref{table-prescriptions} fall within this case, and Higgs inflation with vacuum metastability was discussed in Ref. \cite{Bezrukov:2014ipa} with prescription I and the jump with a sufficiently large negative value of (\ref{Jump}) assumed.
In any case,
\be
{\bm \lambda}_{\rm SI}^{\xi}(\mu_\star ; \mu) \equiv {\bm \lambda}_{\rm ch}(\infty;\mu_\star ; \mu)= {\bm \lambda}_{\rm ch}(\mu_\star;\infty ; \mu)  \label{lambda_SI^xi}
\ee
gives the asymptotic value of ${\bm \lambda}_{\rm aSI}^{\xi}$ (see footnote \ref{infinity-mu_star} on page \pageref{infinity-mu_star}).

\

Let us comment on higher-loop corrections.
In our model, all two-loop vacuum diagrams and higher-loop ones involve the scalar fluctuations to be suppressed for $\ul{h} \gg \mu_{\rm km}$ since each scalar propagator comes with $(1+\ul{h}^2 /\mu_{\rm km}^2)^{-1}$ originating from (\ref{field-space-metric}).
In this sense, all higher-loop contributions are regarded as corrections to ${\bm \lambda}_{\rm km}$, ${\bm \lambda}_{\vs}$ and ${\bm \lambda}_{\wt{\star}}$ and consistent with asymptotic scale invariance despite of $\ln z$ as mentioned above.
When we identify our fermion as the top quark, it interacts with the SM gauge bosons to generate higher-loop vacuum diagrams without the fluctuations of the canonically normalized Higgs boson $\chi$.
Those survive for $\ul{h} \gg \mu_{\rm km}$ to give corrections to ${\bm \lambda}_{\rm ch}$.
The masses of the gauge bosons are simply proportional to $\ul{h}/\ul{\Omega}$ in the Einstein frame as is the case with fermion's mass, and then,
the higher-loop corrections give
\be
{\bm \lambda}_{\rm ch}(\ul{h};\mu_\star;\mu) &=& \sum_{l=0}^{\infty}\frac{{\mf B}_{\rm ch}^{(l)}}{l! (4 \pi)^{2l}} \left(\ln \frac{\ul{h}}{\ul{\omega}}\right)^l  \nn %\label{effective-lambda_ch}
\ee
with ${\mf B}^{(l)}_{\rm ch}  \equiv (4 \pi)^{2l}  \left[ \{-\mu \partial_{\mu} \}^{l} {\bm \lambda}_{\rm ch}(\ul{h};\mu_\star;\mu) \right]_{\ul{h}=\ul{\omega}} $
which respects asymptotic scale invariance as (\ref{effective-lambda_aSI}) does.
And the corresponding beta function is nothing but the one following from the chiral EW Lagrangian; see Ref. \cite{Bezrukov:2009db} for a discussion.
Needless to say, ${\bm \lambda}_{\rm km}$, ${\bm \lambda}_{\vs}$ and ${\bm \lambda}_{\wt{\star}}$ also acquire corrections from the gauge bosons.

%%%%%%%%%%%%%%%%%%%%%%%%%%%%%%%%%%%%%%%
%%%%%%%%%%%%%%%%%%%%%%%%%%%%%%%%%%%%%%%
%%%%%%%%%%%%%%%%%%%%%%%%%%%%%%%%%%%%%%%

\section{\label{Cosmological consequences}Cosmological consequences}

In this section,
identifying the scalar field $h$ as the SM Higgs field,
we discuss possible cosmological scenarios with asymptotic scale invariance with $\mu_\star$ as a free parameter.
When a large nonminimal coupling is assumed, we find various shapes of the effective potential and related phenomena.
For a small nonminimal coupling, $\mu_\star$ is the only scale which modifies the effective potential below the Planck scale, and an interesting consideration can be made in association with the SM near-criticality.

%%%%%%%%%%%%%%%%%%%%%%%%%%%%%%%%%%%%%%%
\subsection{\label{Cosmology with large nonminimal coupling}Cosmology with large nonminimal coupling}
Here, we focus on the case with a large nonminimal coupling $\xi \gg 1$ where corrections to the effective potential due to the quantum fluctuation of the Einstein frame metric are suppressed as mentioned at the end of Sec. \ref{From Jordan to Einstein}.
Now we have 
\be
\mu_{\rm km} \ll \mu_{\rm si} \ll M_P  \label{restriction-1}
\ee
and the new physics/strong coupling scale (\ref{field-dependent-cutoff}) is significantly lower than the Planck scale regardless of $\mu_\star$.

Let us define field values $h_\star$ and $h_{\star\star}$ as
\be
&&{\bm \lambda}_{\rm aSI}^{\xi}(h_{\star/\star \star};\infty,\infty,\infty;\mu) = 0
%\nn \\ &&
~~~ {\rm with} ~~~ \partial_{\mu}{\bm \lambda}_{\rm aSI}^{\xi}(h_{\star/\star \star};\infty,\infty,\infty;\mu) \gtrless 0 ~. \nn
\ee
Note that the definition of $h_\star$ here is equivalent to (\ref{h_star}) since $\mu_{\star/{\rm km}/{\rm si}} = \infty$ reproduces the standard result without the nonminimal coupling.
As a phenomenologically interesting case, we assume
\be
h_\star < M_P < h_{\star \star} ~,  \label{restriction-2}
\ee
which is to say, without both the nonminimal coupling and asymptotic scale invariance, the Higgs effective potential becomes negative below the Planck scale and the EW vacuum is metastable with the true vacuum at a certain trans-Planckian field value.

However, as seen in Sec. \ref{lambda stops ``running'' before/after it jumps}, ${\bm \lambda}_{\rm aSI}^{\xi}$ is replaced by ${\bm \lambda}_{\rm ch}$ once the field value becomes larger than $\mu_{\rm km}$.
So it is convenient to introduce field values $h_{\star}^{\rm ch}$ and $h_{\star \star}^{\rm ch}$ by 
\be
{\bm \lambda}_{\rm ch}(h^{\rm ch}_{\star/\star \star};\infty;\mu) = 0 ~~~~ {\rm with} ~~~~ \partial_{\mu}{\bm \lambda}_{\rm ch}(h^{\rm ch}_{\star/\star \star};\infty;\mu) \gtrless 0  ~. \nn
\ee
Note that these field values depend on the UV completion in the sense that how ${\bm \lambda}_{\rm ch}$ behaves depends on the boundary condition specified by $\mf C$ in (\ref{V_1^chi}) which cannot be fixed by the low-energy EW physics.\footnote{
As discussed in Ref. \cite{Bezrukov:2014ipa},
if the top quark is as heavy as $173$ GeV, noncritical Higgs inflation with prescription I requires that $\Delta$ defined in (\ref{Jump}) is larger than ${\cal O}(10^{-2})$.
Note that the field values $h^{\rm ch}_{\star/\star\star}$ also depend on the counterpart of $\mf C$ in the Yukawa interaction with the renormalization group improvement or corrections beyond the one-loop level.
}
In the following, we assume
\be
h_{\star}^{\rm ch} < h_{\star \star}^{\rm ch} \nn
\ee
while the relationship with the Planck scale is unrestricted. 
If ${\bm \lambda}_{\rm ch}(\ul{h};\infty;\mu)$ never crosses zero,
these field values are understood to be infinitely large and replaced by a finite field value $h^{\rm ch}_{c}$ where $ \partial_{\mu}{\bm \lambda}_{\rm ch}(h^{\rm ch}_{c};\infty;\mu) = 0$ holds with ${\bm \lambda}_{\rm ch}(h^{\rm ch}_{c};\infty;\mu) >  0$.

%%%%%%%%%%%%%%%%%%%%%%%%%%%%%%%%%%%%%%%
\subsubsection{\label{Effective potential with xi>>1}Effective potential with $\xi \gg 1$}
Note that, given (\ref{restriction-1}) and (\ref{restriction-2}),
the field value $h_{\star \star}$ is no longer relevant since ${\bm \lambda}_{\rm aSI}^{\xi}(\ul{h};\mu_\star , \mu_{\rm km}, \mu_{\rm si};\mu) \approx {\bm \lambda}_{\rm ch}(\ul{h};\mu_\star;\mu)$ for $\ul{h} \gtrsim \mu_{\rm km}$.
For notational simplicity, let us introduce 
\be
h_{-} \equiv {\rm min}\{h_\star , h^{\rm ch}_\star\} ~~ ,~~~ h_{+} \equiv {\rm max}\{h_\star , h^{\rm ch}_\star\} \nn
\ee
and also
\be
\mu_{+} \equiv {\rm max}\{\mu_\star , \mu_{\rm km} \}  \nn
\ee
in addition to $\mu_{-}$ defined as (\ref{mu_-}).

Depending on the magnitude relationship among $h_\star$, $h^{\rm ch}_{\star}$, $h^{\rm ch}_{\star \star}$, $\mu_\star$ and $\mu_{\rm km}$,
we find various types of the potential shape which can be classified according to the number of zero crossings of ${\bm \lambda}_{\rm aSI}^{\xi} = \wt{\bm V}_{\rm aSI}^{\xi}/(\ul{h}^4 /4 \ul{\Omega}^4)$.
It is straightforward to obtain following conditions for each case, based on (\ref{rough-behavior-1}) and (\ref{rough-behavior-2}):
%%%%%%%%
\begin{figure}[t]
\centering
\includegraphics[width=1 \linewidth, bb=0 0 708 440]{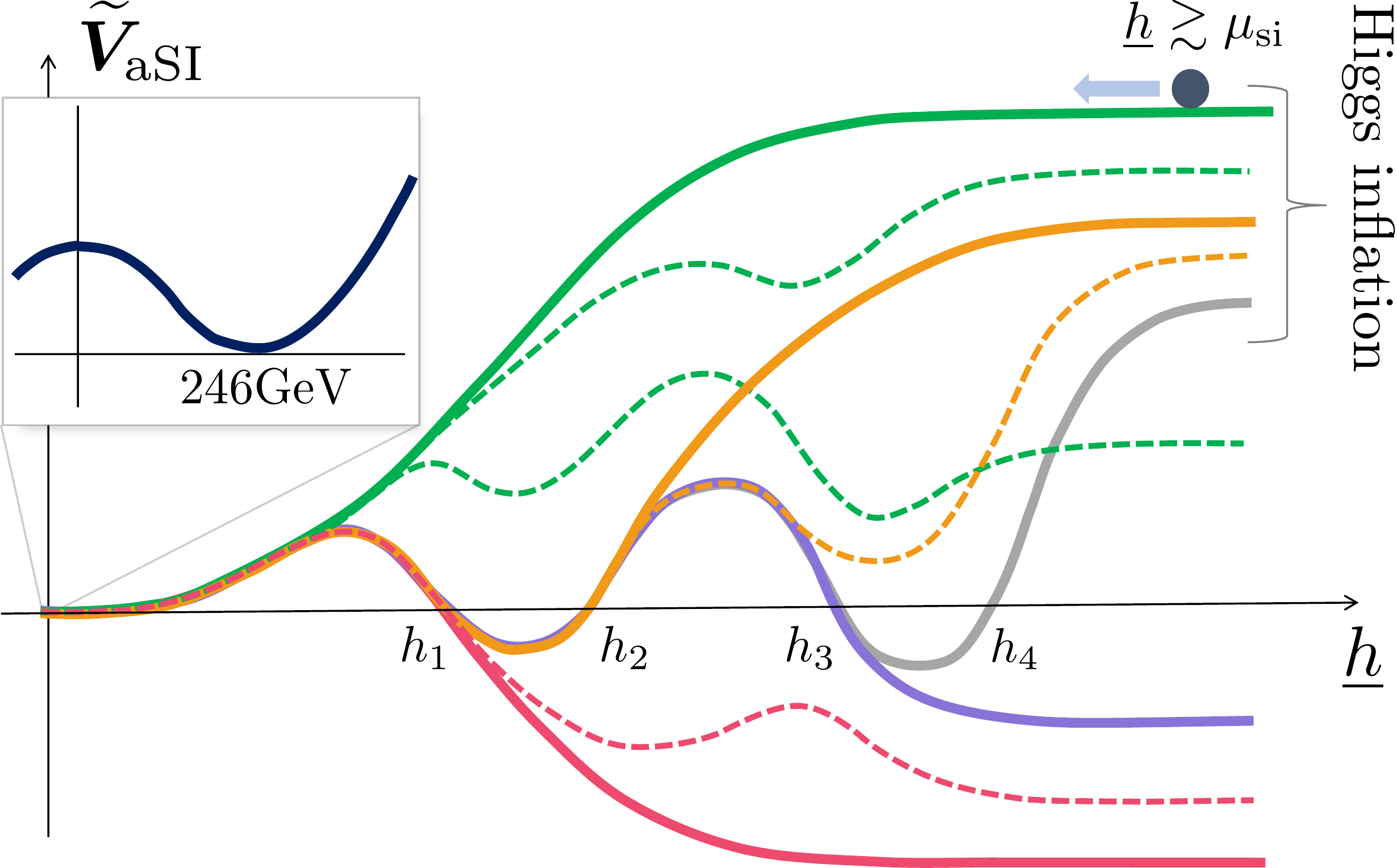}
\caption{\label{Fig:noncritical}Schematic view of various shapes of the effective potential $\wt{\bm V}_{\rm aSI} = {\bm \lambda}_{\rm aSI}^{\xi} \ul{h}^4 / 4 \ul{\Omega}^4$, color coded according to the number of zero crossings.
$\wt{\bm V}_{\rm aSI}$ may wiggle without a change in sign, as depicted with the dashed lines.
Depending on the shape, the Higgs inflation scenario can be realized for $\ul{h} \gtrsim \mu_{\rm si}$. 
}
\end{figure}
%%%%%%%
\begin{itemize}
\item ${\bm \lambda}_{\rm aSI}^{\xi}$ does {\it not} cross zero (the green lines in Fig. \ref{Fig:noncritical}) and absolute stability is achieved when either
\be
\mu_\star < h_{-} \label{stable-1}
\ee
or 
\be
\mu_{\rm km} <h_\star < \mu_\star < h^{\rm ch}_\star \label{stable-2}
\ee
is satisfied.
The first possibility is the simplest one where $\mu_{\rm km}$ does not play any role.
In the second one, $\mu_{\rm km} <h_\star$ prevents ${\bm \lambda}_{\rm aSI}^{\xi}$ from crossing zero.
%%%%%%%
\item ${\bm \lambda}_{\rm aSI}^{\xi}$ crosses zero {\it once} (the red lines in Fig. \ref{Fig:noncritical}) at
\be
h_1 \sim
\left\{ \begin{matrix} h_\star \\  h^{\rm ch}_\star  \end{matrix}  ~~~ {\rm if} ~~~
\begin{matrix}
h_\star <\mu_\star < h^{\rm ch}_\star <\mu_{\rm km} \\ \mu_{\rm km}< h^{\rm ch}_\star  <\mu_\star  < h_\star
\end{matrix} \right.  ~~.\label{crossing-once-2}
\ee
For these two possibilities, $\mu_\star$ is necessarily smaller than $M_P$ because of our assumptions (\ref{restriction-1}) and (\ref{restriction-2}). 
On the other hand, that is not necessary for
\be
h_1 \sim \left\{ \begin{matrix} \mu_{\rm km}  &  & h^{\rm ch}_\star <\mu_{-} < h_\star \\
h_\star  & \quad {\rm if} \quad & h_{+} <\mu_{-} \\
h^{\rm ch}_\star &  & \mu_{\rm km} < h_{-} ~ {\rm and} ~~  h_{+} < \mu_\star 
\end{matrix} \right.  \label{crossing-once}
\ee
with $\mu_\star < h_{\star \star}^{\rm ch}$.
The case with $h_1 \sim \mu_{\rm km}$ corresponds to the example shown in Fig. \ref{Fig:lambda_aSI^xi-1} with the black line.
%%%%%%%
\item ${\bm \lambda}_{\rm aSI}^{\xi}$ crosses zero {\it twice} (the orange lines in Fig. \ref{Fig:noncritical}) at (\ref{crossing-once}) first and subsequently at $h_2 \sim h^{\rm ch}_{\star \star}$ when $\mu_{\star} > h^{\rm ch}_{\star \star}$.
As the other possibility, we have 
\be
\begin{aligned}
(h_1, h_2) \sim (h_\star , \mu_{\rm km} )
~~~~{\rm if} ~~~~ h_\star < \mu_{-} ~~ {\rm and} ~~~ \mu_{+} < h^{\rm ch}_\star  ~.
\end{aligned} 
\label{crossing-twice}
\ee
%%%%%%%
\item ${\bm \lambda}_{\rm aSI}^{\xi}$ crosses zero {\it three times} (the purple line in Fig. \ref{Fig:noncritical}) at  
\be
\begin{aligned}
(h_1,h_2,h_3) \sim (h_\star ,\mu_{\rm km}, h^{\rm ch}_\star )
~~~~{\rm if} ~~~~ h_\star <\mu_{\rm km} < h^{\rm ch}_\star < \mu_\star
\end{aligned}
\label{crossing-three-times}
\ee
with $\mu_\star < h^{\rm ch}_{\star \star}$ which is the case shown in Fig. \ref{Fig:lambda_aSI^xi-2} with the black line.
%%%%%%%%
\item ${\bm \lambda}_{\rm aSI}^{\xi}$ crosses zero {\it four times} (the gray line in Fig. \ref{Fig:noncritical}) at (\ref{crossing-three-times}) and subsequently at $h_4 \sim h^{\rm ch}_{\star \star}$ with $\mu_\star > h^{\rm ch}_{\star \star}$.
\end{itemize}

\

In any case, the asymptotic behavior of the effective potential is understood as follows.
For $\ul{h} \gg \mu_{\rm km}$,
it is simply given by\footnote{\label{threshold-correction}
The scalar field value needs to be such large that $|{\bm \lambda}_{\rm aSI}^{\xi} - {\bm \lambda}_{\rm ch}|\ll |{\bm \lambda}_{\rm ch}|$.
With the field dependence of the $z^2$ factors in (\ref{V_1^chi}) and (\ref{u_vs/star}) taken into account, at $\ul{h}\sim \mu_{\rm si}$ for instance, this inequality is roughly equivalent to
$|\Delta| + |\delta| \ll |{\bm \lambda}_{\rm ch}| \xi^2$.
This kind of noninstantaneous threshold correction impacts on the critical Higgs inflation scenario where the nonminimal coupling is relatively small \cite{Bezrukov:2017dyv}; otherwise, it does not affect the cosmic microwave background (CMB) spectrum \cite{Fumagalli:2016lls}.}
\be
\wt{\bm V}_{\rm aSI} \approx \frac{{\bm \lambda}_{\rm ch}(\ul{h};\mu_\star;\mu)}{4} \times \frac{\ul{h}^4}{\ul{\Omega}^4} ~, \label{V-for-h>mu_km} 
\ee
where
\be
{\bm \lambda}_{\rm ch}(\ul{h};\mu_\star;\mu) &=& \sum_{l=0}^{\infty}\frac{{\mf B}_{{\rm ch},\infty}^{(l)}}{l! (4 \pi)^{2l} 2^l} \left(\ln \frac{\ul{h}^2}{\mu_\star^2 +\ul{h}^2}\right)^l  \nn 
\ee
with ${\mf B}^{(l)}_{{\rm ch},\infty}  \equiv (4 \pi)^{2l}   \{-\mu \partial_{\mu} \}^{l} {\bm \lambda}_{\rm ch}(\mu_\star;\infty;\mu)$.
In the large field regime (\ref{large-field}), the effective potential approaches the asymptotic value as 
\be
\wt{\bm V}_{\rm aSI} &\approx & 3 M_P^2 ~\wt{H}_{\infty}^2 \label{inflationary-plateau}\\
&\times &\left[ 1 - \frac{\mu_{\star}^2}{2 \ul{h}^2} \left( \frac{4 \mu_{\rm si}^2}{\mu_{\star}^2} + \frac{{\mf B}^{(1)}_{{\rm ch},\infty}}{(4 \pi)^2 {\mf B}^{(0)}_{{\rm ch},\infty}} \right)  +\cdots\right] ~, \nn
\ee
where
\be
\wt{H}_{\infty}
\equiv \sqrt{\frac{{\mf B}^{(0)}_{{\rm ch},\infty}}{12}} \times \frac{\mu_{\rm si}^2}{M_P} \approx \sqrt{\frac{{\mf B}^{(0)}_{{\rm ch},\infty}}{2}} \times \mu_{\rm km} \label{H_infinity}
\ee
is, if ${\mf B}^{(0)}_{{\rm ch},\infty} > 0$, the Hubble rate of the exponential expansion that occurs on the asymptotic plateau and the combination in the parentheses in (\ref{inflationary-plateau}) 
needs to be positive for the Higgs field to roll down classically from the plateau.
The ellipsis represents terms of higher negative powers of $\ul{h}^{2}$.
As a special case, if ${\mf B}^{(0)}_{{\rm ch},\infty} = 0$, the effective potential asymptotically vanishes as $\propto \exp (- \sqrt{2/3}~\ul{\chi}/M_P)$.

%%%%%%%%%%%%%%%%%%%%%%%%%%%%%%%%%%%%%%%
\subsubsection{\label{Implications for cosmology}Implications for cosmology}
Phenomenologically, the number of zero crossings is not of primary interest.
Rather, based on the above, the following discussion emerges.

\

1. {\it The vacuum stability issue}. ---
When $\mu_{-} < M_{P}$ as assumed here,
the effective potential for $\ul{h} \gtrsim \mu_{-}$ can be significantly different from the one that we expect from the low-energy physics.
In other words, it depends on the UV completion, and so does the lifetime of our electroweak vacuum.
Therefore, the vacuum stability issue needs to be revisited with assumptions about the UV completion; see Refs. \cite{Branchina:2013jra,Branchina:2014rva} for discussions about the impact of the higher-dimensional operators such as $h^6/M_P^2$ and $h^8/M_P^4$.
It should be noted that, although our effective potential can be expanded as the sum of an infinite series of such higher-dimensional operators, the underlying assumption is different.

The simplest scenario to realize the {\it absolute} stability with the condition (\ref{stable-1}) is such that all the nonpolynomial operators are radiatively generated to be loop suppressed, as implied by the tree level Lagrangian (\ref{Einsein-aSI-L_HY[0]}),
and the scale $\mu_\star$ is sufficiently small that ${\bm \lambda}_{\rm ch}$ stops ``running'' while being positive and of the order of the tree level coupling 
to dominate ${\bm \lambda}_{\rm aSI}^{\xi}$ all the time.
Then, the effective potential monotonically increases as the green solid line in Fig. \ref{Fig:noncritical}.
Note that, in this scenario, the ``jump'' at $\ul{h}\sim \mu_{\rm km}$ does not play any role.\footnote{Here, we should refer readers to other approaches for stabilizing the EW vacuum and making Higgs inflation possible.
In Ref. \cite{DiVita:2015bha}, it was assumed that all the SM particles have nonminimal derivative couplings to gravity without new degrees of freedom.
In Ref. \cite{Xianyu:2014eba}, it was speculated that asymptotic safety of gravity is achieved below the instability scale.}

\

2. {\it Higgs inflation}. ---
If the number of zero crossings is even, the asymptotic value of the effective potential is positive.
And if nothing prevents the Higgs field from rolling down towards the EW vacuum from the plateau for $\ul{h} \gtrsim \mu_{\rm si}$, Higgs inflation is possible.
A metastable case with $\mu_\star = \mu_{\rm si}$ is investigated in Ref. \cite{Bezrukov:2014ipa}, and it was shown that, after inflation, the thermal potential can trap the Higgs field at the origin to avoid the collapse of the Universe \cite{Felder:2002jk}.
This possibility corresponds to (\ref{crossing-twice}),
depicted with the orange solid line in Fig. \ref{Fig:noncritical}.

With our prescription, $\mu_\star$ is the new parameter to control the asymptotic value of ${\bm \lambda}_{\rm ch}$ in (\ref{V-for-h>mu_km}) which is given by ${\mf B}^{(0)}_{{\rm ch},\infty}$.
Especially for $\mu_\star \ll \mu_{\rm si}$, the field dependence of ${\bm \lambda}_{\rm ch}$ is totally negligible,
and the inflationary predictions are the same as in the so-called universal/noncritical regime:
once the scalar power spectrum amplitude is accounted for by ${\mf B}^{(0)}_{{\rm ch},\infty} /\xi^2 \simeq 4 \times 10^{-10}$, the other inflationary observables, the spectral tilt $n_s$ and the tensor-to-scalar ratio $r$, are insensitive to the individual value of ${\mf B}^{(0)}_{{\rm ch},\infty}$ with the large nonminimal coupling $\xi$; see e.g. Ref. \cite{Bezrukov:2013fca}.
Although $\mu_\star$ can be tuned to make ${\mf B}^{(0)}_{{\rm ch},\infty}$ such small that a relatively small value of $\xi$ is required, then the noninstantaneous threshold correction (see footnote \ref{threshold-correction}) cannot be neglected.

\begin{figure}[t]
\centering
\includegraphics[width=1 \linewidth, bb=0 0 639 366]{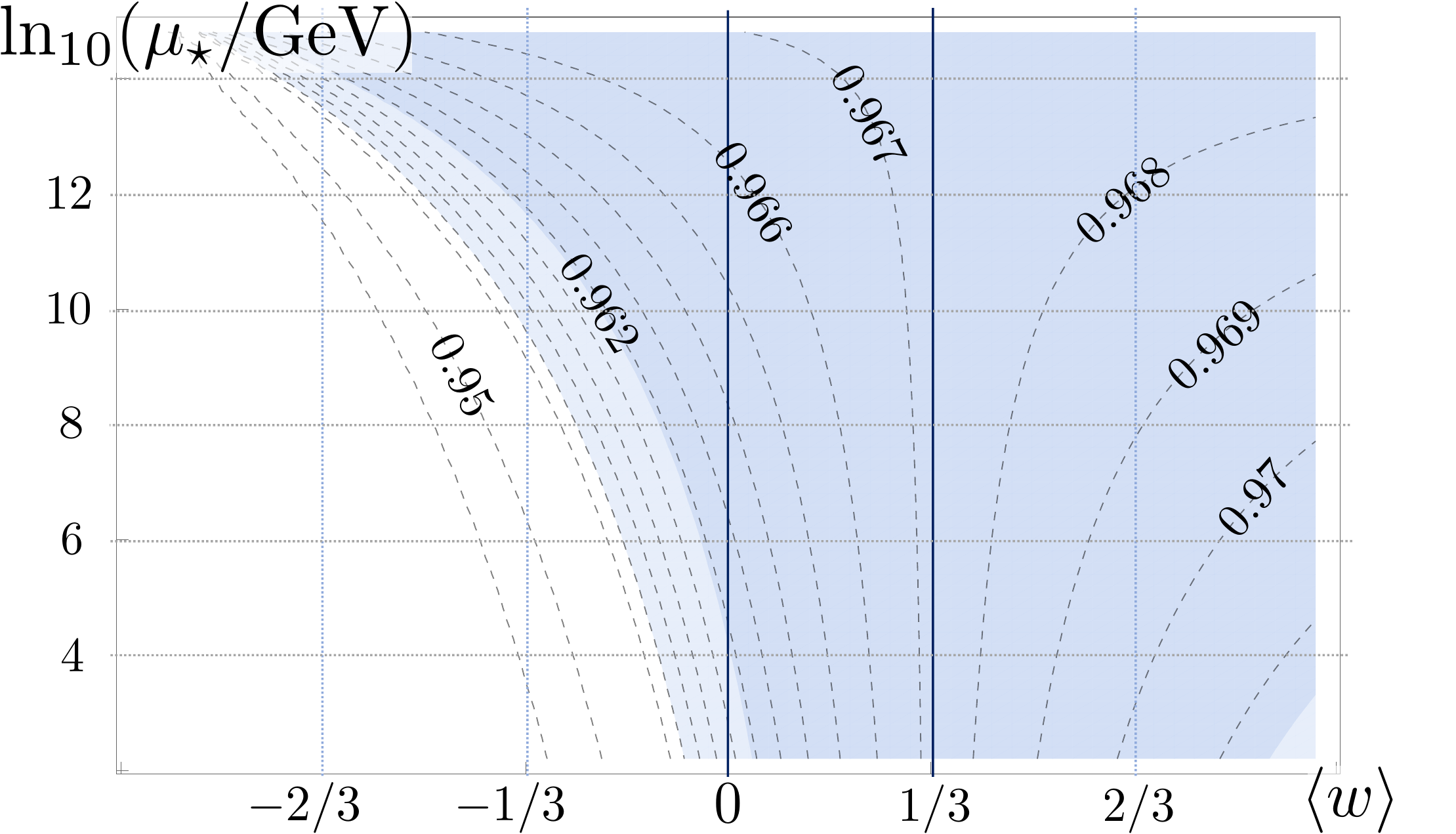}
\caption{\label{Fig:ns-w} Contour plot of the spectral tilt $n_s$ in the universal/noncritical regime as a function of the $e$-folds-averaged EOS $\langle w \rangle$ in the high-energy phase and the scale $\mu_\star < \mu_{\rm km}$ below which the low-energy theory is reliable. The blue (darker) shaded region corresponds to the 2$\sigma$ (1$\sigma$) limit of $n_s$ with $r \simeq 4 \times 10^{-3}$ in the Planck 2018 results \cite{Akrami:2018odb}.}
\end{figure}
Here we note that, with $\mu_{\star} < \mu_{\rm km}$, the tree unitarity violation scale $\wt{\Lambda}_-$ given by (\ref{field-dependent-cutoff}) becomes lower than the one with prescription I for the small field regime $\ul{h} \lesssim \mu_{\rm km}$ and the reheating temperature after Higgs inflation can be higher than it \cite{Bezrukov:2008ut,GarciaBellido:2008ab,Repond:2016sol,Ema:2016dny}.
However, on the other hand, $\wt{\Lambda}_-$ during Higgs inflation remains higher than the momentum scale of the quantum fluctuations of the Higgs field induced by the inflationary expansion with the Hubble rate (\ref{H_infinity}).
Therefore, once we assume that, even above $\wt{\Lambda}_-$, the classical picture of the Friedmann-Lema\^\i tre-Robertson-Walker universe with the super horizon scale perturbation still holds, the effects of the unknown high-energy phase on the inflationary observables are simply parametrized by the effective equation of state (EOS) during the unknown phase.
So let us introduce $\langle w \rangle$ as the $e$-folds-averaged EOS for the period starting at the end of Higgs inflation until the ``phase transition'' after which our low-energy description works well. Identifying $\mu_{-} = \mu_\star$ as the temperature right after the transition, 
we can evaluate the spectral tilt $n_s$ at the pivot scale in the same manner as in Refs. \cite{Bezrukov:2007ep,Bezrukov:2008ut} to find Fig. \ref{Fig:ns-w}.

To the contrary, $\mu_\star > \mu_{\rm si}$ interpolates between prescriptions I and II, each of which is well investigated to be compared against each other \cite{DeSimone:2008ei,Bezrukov:2008ej,Bezrukov:2009db,Allison:2013uaa} in the standard Higgs inflation scenario.\footnote{In Ref. \cite{Hamada:2016onh}, a possibility that prescription II with modified kinetic terms of the SM fields mimics prescription I was mentioned.}
Here, let us mention another possibility where the effective potential is negative at $\ul{h} \sim \mu_{\rm si}$ and crosses zero at $\ul{h} \sim h_{\star \star}^{\rm ch} < \mu_{\star}$ to become positive thanks to the ``running'' of ${\bm \lambda}_{\rm ch} (\ul{h};\mu_\star ; \mu)$
linearly depending on the canonically normalized field \cite{Kannike:2015kda} until $\ul{h}$ reaches $\mu_\star$.
An inflationary period is realized there while the subsequent evolution needs to be carefully studied to obtain the conditions for the Higgs field to end up oscillating around the origin.

\

3. {\it Exotic scenarios with a wiggling effective potential}. ---
If ${\bm \lambda}_{\rm aSI}^{\xi}$ becomes almost vanishing without crossing zero, the effective potential wiggles staying positive or negative, as shown in Fig. \ref{Fig:noncritical} with the dashed lines.
And when it wiggles, the field value at each extremum roughly corresponds to one of $h_\star$, $\mu_{\rm km}$ and $h^{\rm ch}_{\star/\star \star / c}$.

Assume that the Higgs field comes to a wiggle from the inflationary plateau for $\ul{h} \gtrsim \mu_{\rm si}$.
If it almost stops but eventually overcomes the local maximum,
a large-amplitude peak of the density fluctuation is generated there to end up in primordial black holes \cite{Ivanov:1994pa,Yokoyama:1998pt}.
If the nonminimal coupling is relatively small and the inflection point is around $h^{\rm ch}_c \sim M_P$, the so-called critical Higgs inflation scenario \cite{Hamada:2014iga,Bezrukov:2014bra} is realized where primordial black holes can be produced with a relatively large tensor-to-scalar ratio \cite{Garcia-Bellido:2017mdw,Ezquiaga:2017fvi} (see, however Ref. \cite{Bezrukov:2017dyv}).
Even if the local maximum is too high to be overcome, the Higgs field can be the inflaton initially placed at the hilltop \cite{Enckell:2018kkc}.

Even separated from the inflationary period, the wiggling Higgs potential can affect the (p)reheating dynamics and the subsequent thermal history
with possible impacts of the operators suppressed by the Higgs dependent scale on baryogenesis and dark matter production \cite{Bezrukov:2011sz}.
%%%%%%%%%%%%%%%%%%%%%%%%%%%%%%%%%%%%%%%

\subsection{\label{Inflation without nonminimal coupling?}Inflation without nonminimal coupling?}
For completeness,
here we assume $\xi = 0$ as the small nonminimal coupling limit for which $\mu_{{\rm km}/{\rm si}}\to \infty$. 
However, the scale $\mu_\star$ can still be smaller than the Planck scale to affect the Higgs effective potential for $\ul{h} \gtrsim \mu_\star$.
Considering the tree unitarity violation scale (\ref{tree-unitarity-violation-scale-for-xi<1}), we restrict ourselves to $\ul{h} < h_{\rm tuv}$ with $h_{\rm tuv} \sim G_N^{-1/2}$ being a few times larger than $\Lambda_P$ for the SM couplings near the Planck scale.

Without any new degrees of freedom, the Higgs field needs to play the role of the inflaton.
However, at least with the ``standard'' prescription, it is known that the Higgs field cannot be responsible for the observed CMB fluctuation even when the radiative plateau is realized \cite{Isidori:2007vm}.
In the following, we ask if there could be any possibilities of having Higgs inflation without the large nonminimal coupling, opened up by the aSI prescription with $\mu_\star$ as a free parameter.

\

As seen in Sec. \ref{lambda stops ``running''}, the effective potential behaves as if at the tree level (\ref{SI-regime-potential}) for $\ul{h} \gtrsim \mu_\star$ and cannot be consistent with the smallness of the tensor-to-scalar ratio {\it if} the asymptotic value (\ref{lambda_SI}) is finite.
In other words, the scale $\mu_\star$ needs to be tuned so that the asymptotic value vanishes.

An interesting scenario that can possibly be realized within the SM is such that the Higgs quartic coupling vanishes at the same time as its beta function does \cite{Froggatt:1995rt,Bezrukov:2012sa,Degrassi:2012ry,Buttazzo:2013uya,Bednyakov:2015sca}.
At the effective potential level computed with the ``standard'' prescription, this leads to the ``Planck vacuum,'' the local minimum at $h_c \sim M_P$ with vanishing potential energy:
\be
{\bm \lambda}(h_c;\mu) = 0 ~~~ {\rm and }~~~  
 \partial_\mu {\bm \lambda}(h_c;\mu) = 0 ~.\label{SM-criticality}
\ee
From the view point of the low-energy physics, this criticality condition is translated into the condition satisfied by the coupling constants, especially the top Yukawa and the Higgs quartic couplings at the EW scale. 

Let us remind that the field dependent quantity ${\bm \lambda}(\ul{h};\mu)$ is related to ${\bm \lambda}_{[0]}(\ul{h};\mu_\star ;\mu)$ obtained with the aSI prescription as (\ref{effective-lambda_standard}) and, by definition, these are indistinguishable when the field value $\ul{h}$ is much smaller than $\mu_\star$, say at the EW vacuum.
Thus, even if the couplings at the EW scale satisfy the condition corresponding to (\ref{SM-criticality}), the effective potential does not necessarily have the Planck vacuum.
If $\mu_{\star}$ is tuned to coincide with $h_c$ so that ${\bm \lambda}_{[0]}$ vanishes asymptotically,  the condition (\ref{SM-criticality}) turns out to be equivalent to
\be
{\mf B}_{[0],\infty}^{(0)} =0 ~~~ {\rm and } ~~~ {\mf B}_{[0],\infty}^{(1)} = 0  \nn
\ee
via the relation in footnote \ref{infinity-mu_star} on page \pageref{infinity-mu_star}.
Then, expanded with respect to $h_c^2 /\ul{h}^2 < 1$, the field dependent quantity ${\bm \lambda}_{[0]}$ asymptotically vanishes as $\sim h_c^4 / \ul{h}^4$ which, multiplied by $\ul{h}^4$, brings us
\be
{\bm V}_{\rm aSI} \approx \frac{{\mf B}_{[0],\infty}^{(2)} h_c^4}{32 (4 \pi)^4} \left[ 1 - \frac{h_c^2}{\ul{h}^2} + {\cal O}\left(\frac{h_c^4}{\ul{h}^4}\right) \right] ~, \label{flattened-Higgs-potential}
\ee
where only the leading order two-loop part has been kept.
This asymptotic flatness itself is suitable for inflation.\footnote{If ${\mf B}_{[0],\infty}^{(0)} =0$ but with ${\mf B}_{[0],\infty}^{(1)} \ne 0$ corresponding to the case that the Higgs quartic coupling crosses zero, one obtains an asymptotically quadratic potential.}
And with the maximum temperature after inflation bounded from above as $T_{\rm max} <{\bm V}_{\rm aSI}^{1/4} < \mu_\star$, no tree unitarity violation is expected throughout the whole history of the Universe.
However, unfortunately the SM prediction ${\mf B}_{[0],\infty}^{(2)} \sim 1$ together with $h_c \sim M_P$ ends up in too large an amplitude of the scalar power spectrum, roughly 4 orders of magnitude larger than the observed value $A_s \simeq 2.1\times 10^{-9}$.

\

However, there is a logical possibility that the terms neglected above such as the $k\ne 0$ contributions (\ref{lambda_star}) from the nonpolynomial terms and the Planck-suppressed corrections due to the metric fluctuations mentioned at the end of Sec. \ref{From Jordan to Einstein} also contribute to flatten the effective potential, making the inflationary scale lower.
And if the truncation of higher powers of $h_c^2/\ul{h}^2$ in the expansion like (\ref{flattened-Higgs-potential}) is justified, the inflationary observables are well approximated by the ones following from the Coulomb-type potential discussed in the context of the brane inflationary scenario; see Ref. \cite{GarciaBellido:2001ky} and references therein.
If $\mu_\star \sim 10^{17}$GeV flattens the potential, the Higgs field value at the horizon crossing turns out to be typically one order of magnitude smaller than $h_{\rm tuv}$ so that relying on the perturbatively computed effective potential is justified.\footnote{See Ref. \cite{Hamada:2013mya} for another approach assuming a flattened Higgs potential for $\ul{h} \gtrsim 10^{17}$GeV.}

%%%%%%%%%%%%%%%%%%%%%%%%%%%%%%%%%%%%%%%
%%%%%%%%%%%%%%%%%%%%%%%%%%%%%%%%%%%%%%%
%%%%%%%%%%%%%%%%%%%%%%%%%%%%%%%%%%%%%%%

\section{\label{Summary}Summary}

In this work, we introduced the asymptotically scale-invariant (aSI) Standard Model.
Quantum scale invariance in the large Higgs field limit is to be understood as a reflection of the hypothetical scale-invariant nature of the UV completion.

The aSI model is regarded as a low-energy ``effective'' theory in the sense that tree unitarity is violated at a certain energy scale $\Lambda$ above which the perturbative analysis of the scattering amplitudes  becomes unreliable and the theory enters into the strong coupling regime.
By construction, the scale $\Lambda$ increases with the Higgs background value $\ul{h}$,
by virtue of which the Higgs effective potential can be perturbatively computed even for large Higgs field values.
In this large field regime, while the theory still remains in the perturbative regime, the effective potential of the aSI Standard Model can be totally different from that of the canonical SM.

One of the consequences of asymptotic scale invariance is that, even with the current central experimental value of the top quark mass, the EW vacuum can be absolutely stable and successful Higgs inflation may be possible.

\

As introduced in Sec. \ref{Asymptotic scale invariance},
the aSI prescription assumes that the normalization point $\omega$ depends on the Higgs field as $\omega^2 \propto \mu_\star^2 + h^2$ with a mass parameter $\mu_\star$.
In the large field regime $h \gg \mu_\star$, the Higgs field value itself plays the role of the reference scale: the physical significance is given not to the absolute values of various scales but to their ratios to the Higgs field value.
This class of theories turns out to be nonrenormalizable requiring infinitely many counterterms which are suppressed by negative powers of $\Lambda_\star = \mu_\star^2 + h^2$ to be consistent with asymptotic scale invariance.
And then the strong coupling scale is identified as the background value of $\Lambda_\star$.
As explicitly shown in Sec. \ref{lambda stops ``running''},
the effective potential ${\bm V}_{\rm aSI} = {\bm \lambda}_{\rm aSI} \ul{h}^4 /4$, safely computed in the perturbative regime, behaves as $\propto \ul{h}^4$ for $\ul{h} \gtrsim \mu_\star$.
In other words, the field dependent quantity ${\bm \lambda}_{\rm aSI}$ stops ``running''.
While the explicit computation was done in the Higgs-Yukawa model, the result can be straightforwardly applied to the SM.
Then, sufficiently small $\mu_\star$ can render the EW vacuum absolutely stable.

\

We extended the model with the nonminimal coupling to gravity $\xi$ in Sec. \ref{Asymptotic scale invariance with nonminimal coupling to gravity}.
The aSI prescription was introduced in the Jordan frame.
We implemented the Weyl transformation of the spacetime metric to find the normalization point in the Einstein frame $\wt{\omega} = \omega /( 1 + h^2/\mu_{\rm si}^2)^{1/2}$, where $\mu_{\rm si}^2 = M_P^2 /\xi$.
The two conventional choices, prescriptions I and II, are $\mu_\star = \mu_{\rm si}$ and $\mu_\star = \infty$, respectively.
The model becomes shift symmetric in the large field regime $h \gg {\rm max}\{ \mu_\star , \mu_{\rm si}\}$ and, again, nonrenormalizable because of the field dependence of $\wt{\omega}$ as well as the noncanonical kinetic term with ${\cal G}_h$ which reflects the fact that the Higgs-graviton mixing in the Jordan frame becomes significant when $\ul{h}$ exceeds the scale $\mu_{\rm km} = M_P /(6 \xi^2 + \xi)^{1/2}$.
With the SM gauge interactions, the associated strong coupling scale in the Einstein frame is identified as $\wt{\Lambda}_- = ( \mu_{-}^2 + \ul{h}^2)^{1/2} / ( 1 + \ul{h}^2/\mu_{\rm si}^2)^{1/2}$.
Nevertheless, the effective potential $\wt{\bm V}_{\rm aSI} = {\bm \lambda}_{\rm aSI}^{\xi} \ul{h}^4 /4 ( 1 + \ul{h}^2/\mu_{\rm si}^2)^2$ can be computed perturbatively, and we found in Sec. \ref{lambda stops ``running'' before/after it jumps} that ${\bm \lambda}^{\xi}_{\rm aSI}$ loses its field dependence in the shift symmetric large field regime.
If the mass parameter $\mu_\star$ is smaller (larger) than the scale $\mu_{\rm km}$ of the Higgs-graviton mixing,
${\bm \lambda}_{\rm aSI}^{\xi}$ stops ``running''  before (after) 
the ``jump'' due to the threshold correction. 

\

Depending on the value of $\mu_\star$ and $\xi$,
we found various shapes of the Higgs effective potential in Sec. \ref{Cosmological consequences}.
For $\xi \gg 1$, there are up to three scales $\mu_\star$, $\mu_{\rm km}$ and $\mu_{\rm si}$ below the Planck scale in the Einstein frame to make it possible to have Higgs inflation with (meta)stability. One can also discuss some consequences of the possible wiggling behavior of the effective potential, such as primordial black hole production and effects on the (p)reheating dynamics. 
Motivated by the SM criticality, we also made a speculation that, even without the nonminimal coupling, the Higgs field could play the role of the inflaton with the scale $\mu_\star$ tuned so that ${\bm \lambda}_{\rm aSI}$ asymptotically vanishes.

\

Assuming the smallness of the Higgs boson mass, we simply omitted it since we worked mostly in the large field regime.
Also, we just assumed a vanishing cosmological constant.
In order to address the hierarchy problem, the aSI model can be promoted to the exactly SI one \cite{Shaposhnikov:2008xb,Shaposhnikov:2008xi} with a new scalar field $\phi$ as mentioned in Sec. \ref{Exactly scale-invariant prescription}.
The discussion in this work is straightforwardly applied to it once the large field limit $h \gg \mu_\star$ is identified as $h \gg  \Xi^{-1/2} \phi$, more properly, the large ``angle'' limit approaching the $h$ axis in the two-dimensional scalar field space.

\ 

Lastly, the most important is to explore the UV completion of the aSI Standard Model.
One possible direction is to pursue the asymptotic safety scenario of gravity \cite{Weinberg:1980gg,Reuter:1996cp} in which quantum scale invariance is achieved at the nontrivial UV fixed point due to the quantum fluctuation of the metric.
Especially, the asymptotically vanishing behavior of ${\bm \lambda}_{\rm aSI}$ assumed in Sec. \ref{Inflation without nonminimal coupling?} could be discussed in connection with the maximally symmetric fixed point \cite{Eichhorn:2017eht} where Higgs shift symmetry is realized.

%%%%%%%%%%%%%%%%%%%%%%%%%%%%%%%%%%%%%%%

\section*{Acknowledgments}
We thank
Fedor Bezrukov and
Sander Mooij for helpful discussions.
This work was supported by the European Research Council (ERC-AdG-2015) Grant No. 694896.
The work of M.S. was supported partially by the Swiss National Science Foundation.
The work of K.S. was supported by a Japan Society for the Promotion of Science (JSPS) postdoctoral fellowship for research abroad.

%%%%%%%%%%%%%%%%%%%%%%%%%%%%%%%%%%%%%%
%%%%%%%%%%%%%%%%%%%%%%%%%%%%%%%%%%%%%%%
%%%%%%%%%%%%%%%%%%%%%%%%%%%%%%%%%%%%%%%

\appendix

\section{Computing the effective potential}
Here, let us provide some details of the perturbative computation of the effective potential and its two-loop level result taken into account in Fig. \ref{fig1}.
\subsection{\label{Differences from the ``standard'' computation}Differences from the ``standard'' computation}
To remove all divergences in the perturbative computation, besides the couplings constants $\lambda$ and $y$, the dynamical fields $h$ and $f$ also need to be regarded as the bare fields $h_{\rm b} = Z_h^{1/2}h$ and $f_{\rm b} = Z_f^{1/2} f$ with the dimensionless factors $Z_{h/f}$.
While this is mentioned below (\ref{bare-couplings}),
the subscripts ``b'' are suppressed thereafter for notational simplicity.
And in general, the ``hidden'' parameter $\mu_\star$ in the normalization point $\omega$ is also to be bare as
\be
\mu_{\star {\rm b}} =  Z_{\star}^{1/2} \mu_{\star} \ . \nn
\ee
These dimensionless factors are expanded in the same manner as the couplings
\be
Z_{h/f/\star}= 1 + \sum_{i=0}^{\infty} \frac{{\sf C}^{h/f/\star}_{(i)}}{\ve^i} ~, \nn
\ee
and the coefficients ${\sf C}_{(i\geq 1)}$ are fixed for canceling the divergences.
With the nonpolynomial operators introduced,
the scalar self-interaction term, for instance, is now written as
\be
\hat{W} &=&  \sum_{k=0}^{\infty} \frac{\hat{\lambda}_{[k]}^{\rm aSI}}{4} \frac{h_{\rm b}^{4+2k}}{(\mu_{\star{\rm b}}^2 + h_{\rm b}^2)^{k}} \nn \\
&=& \frac{h_{\rm b}^4}{4}\sum_{k=0}^{\infty} \hat{\lambda}_{[k]}  \left(\frac{h_{\rm b}^2}{\mu_{\star{\rm b}}^2} \right)^{k} \left( 1 + \frac{h_{\rm b}^2}{\mu_{\star{\rm b}}^2} \right)^{\frac{2 \ve}{1-\ve} - k} ~, \nn 
\ee
where $\hat{\lambda}_{[k]}$ is the field independent dimensionful quantity defined in (\ref{W_HY}).

However, one immediately realizes that there is a redundancy and the coefficients ${\sf C}_{(i\geq 1)}$ are not uniquely fixed.
This becomes obvious when $\hat{W}$ is expanded with respect to $h_{\rm b}^2 / \mu_{\star{\rm b}}^2$ as
\be
\hat{W} &=& \frac{h_{\rm b}^4}{4} \sum_{s=0}^{\infty} \hat{\lambda}_{[s]}'  \left(\frac{h_{\rm b}^2}{\mu_{\star{\rm b}}^2} \right)^{\! s}  \label{standard-like}
\ee
with
\be
\hat{\lambda}_{[s]}' \equiv  \sum_{j = 0}^{s} \hat{\lambda}_{[s-j]} \binom{\frac{2 \ve}{1-\ve} +j-s}{j} ~, \nn
\ee
where 
$\binom{r}{j}= r (r-1) \cdots (r-j+1)/j!$ is the generalized binomial coefficient with $\binom{r}{0}= 1$.
Having the infinitely many couplings $\hat{\lambda}_{[k\geq 1]}$,
one can redefine those to absorb the factor $Z_\star$. 
In other words, we can choose it at our disposal.
A convenient choice we made is $Z_\star = Z_h$ so that $h/\mu_\star = h_{\rm b} / \mu_{\star{\rm b}}$ is scale independent according to which the scale dependences of $\hat{\lambda}_{[k\geq 1]}$ are fixed.

\

Once the redundancy is removed, the effective potential is computed essentially in the same manner as with the ``standard'' prescription but with the evanescent interactions due to the fluctuations of $\omega$.
It can be expanded with respect to $\delta h$ as
\be
\left(\frac{\omega^{2}}{\ul{\omega}^{2}} \right)^{\frac{q \ve}{1-\ve}} =  \sum_{k=0}^{\infty} \sum_{l=0}^{k} \binom{\frac{q \ve}{1-\ve}}{k} \binom{k}{l} \frac{(2 \ul{h})^{k-l} \delta h^{k+l}}{\ul{\Lambda}^{2k}} ~. \nn
\ee
Then, one finds that the coefficients in (\ref{omega-expanded}) are given as,
for the $(2n+1)$th order perturbation,
\be
{\cal E}^{(q)}_{2n+1,2m-1} \equiv \frac{4^m}{2} \binom{\frac{q \ve}{1-\ve}}{n+m} \binom{n+m}{n-m+1} \label{Epsilon-odd}
\ee
with $n+1 \geq m \geq 1$; for the $(2n+2)$th order perturbation,
\be
{\cal E}^{(q)}_{2n+2,2m} \equiv 4^m \binom{\frac{q \ve}{1-\ve}}{n+m+1} \binom{n+m+1}{n-m+1} \label{Epsilon-even}
\ee
with $n+1 \geq m \geq 0$.
Other components are all zero.  

%%%%%%%%%%%%
\subsection{\label{Two-loop level corrections}Two-loop level corrections}
The two-loop vacuum diagrams add to (\ref{Coleman-Weinberg-1-aSI}) the following corrections.
As the ``normal'' part which does not involve the evanescent terms, we have 
\be
{\cal V}_2 (\ul{h},\ul{\omega}) &=& \frac{ \ul{h}^{4}/4}{(4 \pi)^4} \, {\Bigg \{}  b^{\rm L} +  \sum_{x=h,f}  \frac{b^{x}}{2}  \ln \left(\frac{m_{x}^2}{\ul{\omega}^2 e^2} \right) \nn \\ 
&& \qquad \quad + \sum_{x,z=h,f} \frac{B^{xz}}{8}  \ln \left(\frac{m_{x}^2}{\ul{\omega}^2 e} \right)  \ln \left(\frac{m_{z}^2}{\ul{\omega}^2 e} \right)  {\Bigg \}}   \notag
\ee
with
\be
b^{h} &=& 12 \lambda ( 2 y^4 - 18 \lambda^2 - 3 \lambda y^2 ) ~, \nn \\
b^{f} &=& 4 y^4 ( 2 y^2 - 3 \lambda ) ~, \nn \\
B^{hh} &=& 648 \lambda^3 ~, \nn \\
B^{ff} &=& -12 y^2 (y^4 - 6 \lambda y^2 + 6 \lambda^2  ) ~, \nn \\
B^{hf} &=& -72 \lambda y^2 (y^2 - \lambda ) ~, \nn
\ee
\be
b^{\rm L} = 8\sqrt{3} {\Bigg \{ } y^2 \lambda^{1/2} (2 y^2 -3 \lambda)^{3/2} L\left( \sin^{-1} \sqrt{\frac{3\lambda}{2y^2}} \right) - 18 \lambda^3 L\left(\frac{\pi}{6}\right) {\Bigg \} } ~,\nn
\ee
where 
\be
L(\theta) = -\int^{\theta}_0 d\theta' \, \ln \left( 2 \sin \theta' \right) =  \frac{1}{2}\sum_{k=1}^{\infty} \frac{\sin (2 k \theta )}{k^2} \nn
\ee
is the Lobachevsky function and $3\lambda < 2 y^2$ is assumed.
%%%%%%%
In addition, the evanescent contributions are transmitted to the four-dimensional limit as
\be
{\cal U}_2(\ul{h};\ul{\omega}) 
=\frac{\ul{h}^{4}/4}{(4\pi)^4}\sum_{k=1}^{4} \frac{\ul{h}^{2k}}{ \ul{\Lambda}^{2k} } \sum_{x=h,f}\frac{b_{[k]}^x}{2} \ln \left( \frac{m_{x}^2}{\ul{\omega}^2 e^{3/2}} \right)  \nn
\ee
with
\be
b_{[1]}^{h} &=& 18 \lambda \left( 3 y^4 - 7 \lambda  y^2  - 162 \lambda^2  \right) ~, \nn \\
b_{[2]}^{h} &=& -6 \lambda \left( 2 y^4 - 2\lambda y^2  -537 \lambda^2  \right) ~, \nn \\
b_{[3]}^{h} &=& - 1872 \lambda^3 ~, \nn \\
b_{[4]}^{h} &=& 432 \lambda^3 ~, \nn \\
b_{[1]}^{f} &=& 4 y^4  \left( 3 y^2 + 9 \lambda  \right) ~, \nn \\
b_{[2]}^{f} &=& -24 \lambda y^4 ~, \nn
\ee
and
\be
U_{2 \star}(\ul{h}) 
=  \sum_{k=1}^{6} ~ \frac{a_{[k]}}{(4\pi)^{4}} ~ \frac{\ul{h}^{4+2k}}{4 \ul{\Lambda}^{2k}} \nn
\ee
with
\be
a_{[1]} &=& - 4 y^6  + 33 \lambda y^4 /2  + 45 \lambda^2  y^2  +  270 \lambda^3 ~, \nn \\
a_{[2]} &=& 558 \lambda^3 + \lambda y^4 /4 - 3 y^6 /2 + 57 \lambda^2  y^2 /2 ~,  \nn \\
a_{[3]} &=&  - 8157 \lambda^3 /4 - 6 \lambda^2 y^2 + 3 \lambda y^4 ~, \nn \\
a_{[4]} &=&  1467 \lambda^3 - \lambda y^4 ~, \nn \\
a_{[5]} &=& -474 \lambda^3 ~, \nn \\
a_{[6]} &=& 60 \lambda^3 ~. \nn
\ee
Here, the $\overline{\rm MS}$ scheme is used to fix the finite parts ${\sf C}_{[k](0)}^{\lambda}$ in (\ref{W_HY}).

As usual, one can impose $d {\bm V}_{\rm aSI} / d\mu = 0$ to find the beta functions $\beta_{[k]}$ of the couplings $\lambda_{[k]}$.
It turns out that $\beta_{[0]}$ remains the same as with the ``standard'' prescription and
that the leading order contributions to $\beta_{[1 \leq k \leq 4]}$ appear at the two-loop level but the scale dependences for $k\geq 5$ are yet to come as higher-loop corrections.
This is merely because of our assumption (\ref{aSI-L_HY[0]}).
Note also that, if one chooses $Z_\star \neq Z_h$, it leads to $d_{\mu} (h/\mu_\star) \neq 0$ and a different set of $\beta_{[k\geq 1]}$.
However, this simply stems from the difference in how $\mu_{\star {\rm b}}$ and $\hat{\lambda}_{[k\geq 1]}$ are defined in the first place.
In other words, the differences in $\beta_{[k\geq 1]}$ are to be compensated by the difference in the anomalous dimension of $\mu_\star$.

In terms of the so-called technical naturalness, 
the appearance of $\beta_{[1 \leq k \leq 4]}$ at the two-loop level tells us that the typical values of $\lambda_{[1 \leq k \leq 4]}$ are, at least, of the two-loop order.
Those can be consistently taken into account in the finite parts ${\sf C}_{[1 \leq k \leq 4](0)}^{\lambda}$ and then shift the asymptotic value (\ref{lambda_SI}) at the two-loop level.
However, this ambiguity does not affect our discussion.

%\begin{thebibliography}{99}
%
%%\bibitem{Kaplan:2005rr}
%%      D. E. Kaplan and R. Sundrum,
%%      {\it A Symmetry for the cosmological constant},
%%      JHEP {\bf 07}, 042 (2006)
%%      [arXiv:hep-th/0505265].
%%      
%
%                  
%      \end{thebibliography}

\bibliography{aSI-published}

\begin{thebibliography}{100}

\bibitem{Aad:2012tfa}
Georges Aad et~al.
\newblock {Observation of a new particle in the search for the Standard Model
  Higgs boson with the ATLAS detector at the LHC}.
\newblock {\em Phys. Lett.}, B716:1--29, 2012, 1207.7214.

\bibitem{Chatrchyan:2012xdj}
Serguei Chatrchyan et~al.
\newblock {Observation of a new boson at a mass of 125 GeV with the CMS
  experiment at the LHC}.
\newblock {\em Phys. Lett.}, B716:30--61, 2012, 1207.7235.

\bibitem{Maiani:1977cg}
L.~Maiani, G.~Parisi, and R.~Petronzio.
\newblock {Bounds on the Number and Masses of Quarks and Leptons}.
\newblock {\em Nucl. Phys.}, B136:115--124, 1978.

\bibitem{Cabibbo:1979ay}
N.~Cabibbo, L.~Maiani, G.~Parisi, and R.~Petronzio.
\newblock {Bounds on the Fermions and Higgs Boson Masses in Grand Unified
  Theories}.
\newblock {\em Nucl. Phys.}, B158:295--305, 1979.

\bibitem{Lindner:1985uk}
M.~Lindner.
\newblock {Implications of Triviality for the Standard Model}.
\newblock {\em Z. Phys.}, C31:295, 1986.

\bibitem{Bezrukov:2012sa}
Fedor Bezrukov, Mikhail~{\relax Yu}. Kalmykov, Bernd~A. Kniehl, and Mikhail
  Shaposhnikov.
\newblock {Higgs Boson Mass and New Physics}.
\newblock {\em JHEP}, 10:140, 2012, 1205.2893.
\newblock [,275(2012)].

\bibitem{Degrassi:2012ry}
Giuseppe Degrassi, Stefano Di~Vita, Joan Elias-Miro, Jose~R. Espinosa, Gian~F.
  Giudice, Gino Isidori, and Alessandro Strumia.
\newblock {Higgs mass and vacuum stability in the Standard Model at NNLO}.
\newblock {\em JHEP}, 08:098, 2012, 1205.6497.

\bibitem{Buttazzo:2013uya}
Dario Buttazzo, Giuseppe Degrassi, Pier~Paolo Giardino, Gian~F. Giudice,
  Filippo Sala, Alberto Salvio, and Alessandro Strumia.
\newblock {Investigating the near-criticality of the Higgs boson}.
\newblock {\em JHEP}, 12:089, 2013, 1307.3536.

\bibitem{Bednyakov:2015sca}
A.~V. Bednyakov, B.~A. Kniehl, A.~F. Pikelner, and O.~L. Veretin.
\newblock {Stability of the Electroweak Vacuum: Gauge Independence and Advanced
  Precision}.
\newblock {\em Phys. Rev. Lett.}, 115(20):201802, 2015, 1507.08833.

\bibitem{Tanabashi:2018oca}
M.~Tanabashi et~al.
\newblock {Review of Particle Physics}.
\newblock {\em Phys. Rev.}, D98(3):030001, 2018.

\bibitem{Krasnikov:1978pu}
N.~V. Krasnikov.
\newblock {Restriction of the Fermion Mass in Gauge Theories of Weak and
  Electromagnetic Interactions}.
\newblock {\em Yad. Fiz.}, 28:549--551, 1978.

\bibitem{Politzer:1978ic}
H.~David Politzer and Stephen Wolfram.
\newblock {Bounds on Particle Masses in the Weinberg-Salam Model}.
\newblock {\em Phys. Lett.}, 82B:242--246, 1979.
\newblock [Erratum: Phys. Lett.83B,421(1979)].

\bibitem{Hung:1979dn}
Pham~Quang Hung.
\newblock {Vacuum Instability and New Constraints on Fermion Masses}.
\newblock {\em Phys. Rev. Lett.}, 42:873, 1979.

\bibitem{Bezrukov:2014ina}
Fedor Bezrukov and Mikhail Shaposhnikov.
\newblock {Why should we care about the top quark Yukawa coupling?}
\newblock {\em J. Exp. Theor. Phys.}, 120:335--343, 2015, 1411.1923.
\newblock [Zh. Eksp. Teor. Fiz.147,389(2015)].

\bibitem{Farrar:2017eqq}
Glennys~R. Farrar.
\newblock {Stable Sexaquark}.
\newblock 2017, 1708.08951.

\bibitem{Chung:2003fi}
D.~J.~H. Chung, L.~L. Everett, G.~L. Kane, S.~F. King, Joseph~D. Lykken, and
  Lian-Tao Wang.
\newblock {The Soft supersymmetry breaking Lagrangian: Theory and
  applications}.
\newblock {\em Phys. Rept.}, 407:1--203, 2005, hep-ph/0312378.

\bibitem{Feng:2013pwa}
Jonathan~L. Feng.
\newblock {Naturalness and the Status of Supersymmetry}.
\newblock {\em Ann. Rev. Nucl. Part. Sci.}, 63:351--382, 2013, 1302.6587.

\bibitem{Rubakov:2001kp}
V.~A. Rubakov.
\newblock {Large and infinite extra dimensions: An Introduction}.
\newblock {\em Phys. Usp.}, 44:871--893, 2001, hep-ph/0104152.
\newblock [Usp. Fiz. Nauk171,913(2001)].

\bibitem{Davoudiasl:2009cd}
Hooman Davoudiasl, Shrihari Gopalakrishna, Eduardo Ponton, and Jose Santiago.
\newblock {Warped 5-Dimensional Models: Phenomenological Status and
  Experimental Prospects}.
\newblock {\em New J. Phys.}, 12:075011, 2010, 0908.1968.

\bibitem{Perelstein:2005ka}
Maxim Perelstein.
\newblock {Little Higgs models and their phenomenology}.
\newblock {\em Prog. Part. Nucl. Phys.}, 58:247--291, 2007, hep-ph/0512128.

\bibitem{Panico:2015jxa}
Giuliano Panico and Andrea Wulzer.
\newblock {The Composite Nambu-Goldstone Higgs}.
\newblock {\em Lect. Notes Phys.}, 913:pp.1--316, 2016, 1506.01961.

\bibitem{Gildener:1976ai}
Eldad Gildener.
\newblock {Gauge Symmetry Hierarchies}.
\newblock {\em Phys. Rev.}, D14:1667, 1976.

\bibitem{Vissani:1997ys}
Francesco Vissani.
\newblock {Do experiments suggest a hierarchy problem?}
\newblock {\em Phys. Rev.}, D57:7027--7030, 1998, hep-ph/9709409.

\bibitem{Shaposhnikov:2007nj}
Mikhail Shaposhnikov.
\newblock {Is there a new physics between electroweak and Planck scales?}
\newblock In {\em {Astroparticle Physics: Current Issues, 2007 (APCI07)
  Budapest, Hungary, June 21-23, 2007}}, 2007, 0708.3550.

\bibitem{Shaposhnikov:2009pv}
Mikhail Shaposhnikov and Christof Wetterich.
\newblock {Asymptotic safety of gravity and the Higgs boson mass}.
\newblock {\em Phys. Lett.}, B683:196--200, 2010, 0912.0208.

\bibitem{Wetterich:2011aa}
C.~Wetterich.
\newblock {Where to look for solving the gauge hierarchy problem?}
\newblock {\em Phys. Lett.}, B718:573--576, 2012, 1112.2910.

\bibitem{Farina:2013mla}
Marco Farina, Duccio Pappadopulo, and Alessandro Strumia.
\newblock {A modified naturalness principle and its experimental tests}.
\newblock {\em JHEP}, 08:022, 2013, 1303.7244.

\bibitem{Karananas:2017mxm}
Georgios~K. Karananas and Mikhail Shaposhnikov.
\newblock {Gauge coupling unification without leptoquarks}.
\newblock {\em Phys. Lett.}, B771:332--338, 2017, 1703.02964.

\bibitem{Asaka:2005an}
Takehiko Asaka, Steve Blanchet, and Mikhail Shaposhnikov.
\newblock {The nuMSM, dark matter and neutrino masses}.
\newblock {\em Phys. Lett.}, B631:151--156, 2005, hep-ph/0503065.

\bibitem{Asaka:2005pn}
Takehiko Asaka and Mikhail Shaposhnikov.
\newblock {The nuMSM, dark matter and baryon asymmetry of the universe}.
\newblock {\em Phys. Lett.}, B620:17--26, 2005, hep-ph/0505013.

\bibitem{Burgess:2009ea}
C.~P. Burgess, Hyun~Min Lee, and Michael Trott.
\newblock {Power-counting and the Validity of the Classical Approximation
  During Inflation}.
\newblock {\em JHEP}, 09:103, 2009, 0902.4465.

\bibitem{Barbon:2009ya}
J.~L.~F. Barbon and J.~R. Espinosa.
\newblock {On the Naturalness of Higgs Inflation}.
\newblock {\em Phys. Rev.}, D79:081302, 2009, 0903.0355.

\bibitem{Bezrukov:2010jz}
F.~Bezrukov, A.~Magnin, M.~Shaposhnikov, and S.~Sibiryakov.
\newblock {Higgs inflation: consistency and generalisations}.
\newblock {\em JHEP}, 01:016, 2011, 1008.5157.

\bibitem{Cornwall:1974km}
John~M. Cornwall, David~N. Levin, and George Tiktopoulos.
\newblock {Derivation of Gauge Invariance from High-Energy Unitarity Bounds on
  the s Matrix}.
\newblock {\em Phys. Rev.}, D10:1145, 1974.
\newblock [Erratum: Phys. Rev.D11,972(1975)].

\bibitem{Weinberg:1980gg}
Steven Weinberg.
\newblock {ULTRAVIOLET DIVERGENCES IN QUANTUM THEORIES OF GRAVITATION}.
\newblock In {\em General Relativity: An Einstein Centenary Survey}, pages
  790--831. 1980.

\bibitem{Reuter:1996cp}
M.~Reuter.
\newblock {Nonperturbative evolution equation for quantum gravity}.
\newblock {\em Phys. Rev.}, D57:971--985, 1998, hep-th/9605030.

\bibitem{Dvali:2010bf}
Gia Dvali and Cesar Gomez.
\newblock {Self-Completeness of Einstein Gravity}.
\newblock 2010, 1005.3497.

\bibitem{Dvali:2010jz}
Gia Dvali, Gian~F. Giudice, Cesar Gomez, and Alex Kehagias.
\newblock {UV-Completion by Classicalization}.
\newblock {\em JHEP}, 08:108, 2011, 1010.1415.

\bibitem{Aydemir:2012nz}
Ufuk Aydemir, Mohamed~M. Anber, and John~F. Donoghue.
\newblock {Self-healing of unitarity in effective field theories and the onset
  of new physics}.
\newblock {\em Phys. Rev.}, D86:014025, 2012, 1203.5153.

\bibitem{Shaposhnikov:2018xkv}
Mikhail Shaposhnikov and Andrey Shkerin.
\newblock {Conformal symmetry: towards the link between the Fermi and the
  Planck scales}.
\newblock {\em Phys. Lett.}, B783:253--262, 2018, 1803.08907.

\bibitem{Shaposhnikov:2018jag}
Mikhail Shaposhnikov and Andrey Shkerin.
\newblock {Gravity, Scale Invariance and the Hierarchy Problem}.
\newblock {\em JHEP}, 10:024, 2018, 1804.06376.

\bibitem{Bezrukov:2007ep}
Fedor~L. Bezrukov and Mikhail Shaposhnikov.
\newblock {The Standard Model Higgs boson as the inflaton}.
\newblock {\em Phys. Lett.}, B659:703--706, 2008, 0710.3755.

\bibitem{Bardeen:1995kv}
William~A. Bardeen.
\newblock {On naturalness in the standard model}.
\newblock In {\em {Ontake Summer Institute on Particle Physics Ontake Mountain,
  Japan, August 27-September 2, 1995}}, 1995.

\bibitem{Meissner:2006zh}
Krzysztof~A. Meissner and Hermann Nicolai.
\newblock {Conformal Symmetry and the Standard Model}.
\newblock {\em Phys. Lett.}, B648:312--317, 2007, hep-th/0612165.

\bibitem{Foot:2007as}
Robert Foot, Archil Kobakhidze, and Raymond~R. Volkas.
\newblock {Electroweak Higgs as a pseudo-Goldstone boson of broken scale
  invariance}.
\newblock {\em Phys. Lett.}, B655:156--161, 2007, 0704.1165.

\bibitem{Hambye:2007vf}
Thomas Hambye and Michel H.~G. Tytgat.
\newblock {Electroweak symmetry breaking induced by dark matter}.
\newblock {\em Phys. Lett.}, B659:651--655, 2008, 0707.0633.

\bibitem{Iso:2009ss}
Satoshi Iso, Nobuchika Okada, and Yuta Orikasa.
\newblock {Classically conformal $B-L$ extended Standard Model}.
\newblock {\em Phys. Lett.}, B676:81--87, 2009, 0902.4050.

\bibitem{Holthausen:2009uc}
Martin Holthausen, Manfred Lindner, and Michael~A. Schmidt.
\newblock {Radiative Symmetry Breaking of the Minimal Left-Right Symmetric
  Model}.
\newblock {\em Phys. Rev.}, D82:055002, 2010, 0911.0710.

\bibitem{Hur:2011sv}
Taeil Hur and P.~Ko.
\newblock {Scale invariant extension of the standard model with strongly
  interacting hidden sector}.
\newblock {\em Phys. Rev. Lett.}, 106:141802, 2011, 1103.2571.

\bibitem{Karam:2015jta}
Alexandros Karam and Kyriakos Tamvakis.
\newblock {Dark matter and neutrino masses from a scale-invariant multi-Higgs
  portal}.
\newblock {\em Phys. Rev.}, D92(7):075010, 2015, 1508.03031.

\bibitem{Wetterich:1983bi}
C.~Wetterich.
\newblock {Fine Tuning Problem and the Renormalization Group}.
\newblock {\em Phys. Lett.}, 140B:215--222, 1984.

\bibitem{Wetterich:2016uxm}
Christof Wetterich and Masatoshi Yamada.
\newblock {Gauge hierarchy problem in asymptotically safe gravity--the
  resurgence mechanism}.
\newblock {\em Phys. Lett.}, B770:268--271, 2017, 1612.03069.

\bibitem{Wetterich:1987fm}
C.~Wetterich.
\newblock {Cosmology and the Fate of Dilatation Symmetry}.
\newblock {\em Nucl. Phys.}, B302:668--696, 1988, 1711.03844.

\bibitem{Foot:2007iy}
Robert Foot, Archil Kobakhidze, Kristian~L. McDonald, and Raymond~R. Volkas.
\newblock {A Solution to the hierarchy problem from an almost decoupled hidden
  sector within a classically scale invariant theory}.
\newblock {\em Phys. Rev.}, D77:035006, 2008, 0709.2750.

\bibitem{Salvio:2014soa}
Alberto Salvio and Alessandro Strumia.
\newblock {Agravity}.
\newblock {\em JHEP}, 06:080, 2014, 1403.4226.

\bibitem{Einhorn:2014gfa}
Martin~B. Einhorn and D.~R.~Timothy Jones.
\newblock {Naturalness and Dimensional Transmutation in Classically
  Scale-Invariant Gravity}.
\newblock {\em JHEP}, 03:047, 2015, 1410.8513.

\bibitem{Shaposhnikov:2008xi}
Mikhail Shaposhnikov and Daniel Zenhausern.
\newblock {Quantum scale invariance, cosmological constant and hierarchy
  problem}.
\newblock {\em Phys. Lett.}, B671:162--166, 2009, 0809.3406.

\bibitem{Mooij:2018hew}
Sander Mooij, Mikhail Shaposhnikov, and Thibault Voumard.
\newblock {Hidden and explicit quantum scale invariance}.
\newblock {\em Phys. Rev.}, D99(8):085013, 2019, 1812.07946.

\bibitem{tHooft:1972tcz}
Gerard 't~Hooft and M.~J.~G. Veltman.
\newblock {Regularization and Renormalization of Gauge Fields}.
\newblock {\em Nucl. Phys.}, B44:189--213, 1972.

\bibitem{Englert:1976ep}
F.~Englert, C.~Truffin, and R.~Gastmans.
\newblock {Conformal Invariance in Quantum Gravity}.
\newblock {\em Nucl. Phys.}, B117:407--432, 1976.

\bibitem{Shaposhnikov:2008ar}
Mikhail~E. Shaposhnikov and Igor~I. Tkachev.
\newblock {Quantum scale invariance on the lattice}.
\newblock {\em Phys. Lett.}, B675:403--406, 2009, 0811.1967.

\bibitem{Shaposhnikov:2009nk}
M.~E. Shaposhnikov and F.~V. Tkachov.
\newblock {Quantum scale-invariant models as effective field theories}.
\newblock 2009, 0905.4857.

\bibitem{Armillis:2013wya}
Roberta Armillis, Alexander Monin, and Mikhail Shaposhnikov.
\newblock {Spontaneously Broken Conformal Symmetry: Dealing with the Trace
  Anomaly}.
\newblock {\em JHEP}, 10:030, 2013, 1302.5619.

\bibitem{Gretsch:2013ooa}
Frederic Gretsch and Alexander Monin.
\newblock {Perturbative conformal symmetry and dilaton}.
\newblock {\em Phys. Rev.}, D92(4):045036, 2015, 1308.3863.

\bibitem{Tamarit:2013vda}
Carlos Tamarit.
\newblock {Running couplings with a vanishing scale anomaly}.
\newblock {\em JHEP}, 12:098, 2013, 1309.0913.

\bibitem{Shaposhnikov:2008xb}
Mikhail Shaposhnikov and Daniel Zenhausern.
\newblock {Scale invariance, unimodular gravity and dark energy}.
\newblock {\em Phys. Lett.}, B671:187--192, 2009, 0809.3395.

\bibitem{GarciaBellido:2011de}
Juan Garcia-Bellido, Javier Rubio, Mikhail Shaposhnikov, and Daniel Zenhausern.
\newblock {Higgs-Dilaton Cosmology: From the Early to the Late Universe}.
\newblock {\em Phys. Rev.}, D84:123504, 2011, 1107.2163.

\bibitem{Bezrukov:2012hx}
Fedor Bezrukov, Georgios~K. Karananas, Javier Rubio, and Mikhail Shaposhnikov.
\newblock {Higgs-Dilaton Cosmology: an effective field theory approach}.
\newblock {\em Phys. Rev.}, D87(9):096001, 2013, 1212.4148.

\bibitem{Rubio:2014wta}
Javier Rubio and Mikhail Shaposhnikov.
\newblock {Higgs-Dilaton cosmology: Universality versus criticality}.
\newblock {\em Phys. Rev.}, D90:027307, 2014, 1406.5182.

\bibitem{Trashorras:2016azl}
Manuel Trashorras, Savvas Nesseris, and Juan Garcia-Bellido.
\newblock {Cosmological Constraints on Higgs-Dilaton Inflation}.
\newblock {\em Phys. Rev.}, D94(6):063511, 2016, 1604.06760.

\bibitem{Karananas:2016kyt}
Georgios~K. Karananas and Javier Rubio.
\newblock {On the geometrical interpretation of scale-invariant models of
  inflation}.
\newblock {\em Phys. Lett.}, B761:223--228, 2016, 1606.08848.

\bibitem{Shkerin:2016ssc}
A.~Shkerin.
\newblock {Electroweak vacuum stability in the Higgs-Dilaton theory}.
\newblock {\em JHEP}, 05:155, 2017, 1701.02224.

\bibitem{Tokareva:2017nng}
Anna Tokareva.
\newblock {A minimal scale invariant axion solution to the strong CP-problem}.
\newblock {\em Eur. Phys. J.}, C78(5):423, 2018, 1705.10836.

\bibitem{Casas:2017wjh}
Santiago Casas, Martin Pauly, and Javier Rubio.
\newblock {Higgs-dilaton cosmology: An inflation^^e2^^80^^93dark-energy
  connection and forecasts for future galaxy surveys}.
\newblock {\em Phys. Rev.}, D97(4):043520, 2018, 1712.04956.

\bibitem{Ghilencea:2015mza}
D.~M. Ghilencea.
\newblock {Manifestly scale-invariant regularization and quantum effective
  operators}.
\newblock {\em Phys. Rev.}, D93(10):105006, 2016, 1508.00595.

\bibitem{Ghilencea:2016ckm}
D.~M. Ghilencea, Z.~Lalak, and P.~Olszewski.
\newblock {Two-loop scale-invariant scalar potential and quantum effective
  operators}.
\newblock {\em Eur. Phys. J.}, C76(12):656, 2016, 1608.05336.

\bibitem{Ghilencea:2016dsl}
D.~M. Ghilencea, Z.~Lalak, and P.~Olszewski.
\newblock {Standard Model with spontaneously broken quantum scale invariance}.
\newblock {\em Phys. Rev.}, D96(5):055034, 2017, 1612.09120.

\bibitem{Ghilencea:2017yqv}
D.~M. Ghilencea.
\newblock {Quantum implications of a scale invariant regularization}.
\newblock {\em Phys. Rev.}, D97(7):075015, 2018, 1712.06024.

\bibitem{Lalak:2018bow}
Zygmunt Lalak and Pawel Olszewski.
\newblock {Vanishing trace anomaly in flat spacetime}.
\newblock {\em Phys. Rev.}, D98(8):085001, 2018, 1807.09296.

\bibitem{Hamada:2016onh}
Yuta Hamada, Hikaru Kawai, Yukari Nakanishi, and Kin-ya Oda.
\newblock {Meaning of the field dependence of the renormalization scale in
  Higgs inflation}.
\newblock {\em Phys. Rev.}, D95(10):103524, 2017, 1610.05885.

\bibitem{Bezrukov:2014ipa}
Fedor Bezrukov, Javier Rubio, and Mikhail Shaposhnikov.
\newblock {Living beyond the edge: Higgs inflation and vacuum metastability}.
\newblock {\em Phys. Rev.}, D92(8):083512, 2015, 1412.3811.

\bibitem{Dicus:2004rg}
Duane~A. Dicus and Hong-Jian He.
\newblock {Scales of fermion mass generation and electroweak symmetry
  breaking}.
\newblock {\em Phys. Rev.}, D71:093009, 2005, hep-ph/0409131.

\bibitem{Coleman:1973jx}
Sidney~R. Coleman and Erick~J. Weinberg.
\newblock {Radiative Corrections as the Origin of Spontaneous Symmetry
  Breaking}.
\newblock {\em Phys. Rev.}, D7:1888--1910, 1973.

\bibitem{Ford:1992pn}
C.~Ford, I.~Jack, and D.~R.~T. Jones.
\newblock {The Standard model effective potential at two loops}.
\newblock {\em Nucl. Phys.}, B387:373--390, 1992, hep-ph/0111190.
\newblock [Erratum: Nucl. Phys.B504,551(1997)].

\bibitem{Bezrukov:2008ut}
F.~Bezrukov, D.~Gorbunov, and M.~Shaposhnikov.
\newblock {On initial conditions for the Hot Big Bang}.
\newblock {\em JCAP}, 0906:029, 2009, 0812.3622.

\bibitem{Barvinsky:2008ia}
A.~O. Barvinsky, A.~{\relax Yu}. Kamenshchik, and A.~A. Starobinsky.
\newblock {Inflation scenario via the Standard Model Higgs boson and LHC}.
\newblock {\em JCAP}, 0811:021, 2008, 0809.2104.

\bibitem{Ferreira:2016vsc}
Pedro~G. Ferreira, Christopher~T. Hill, and Graham~G. Ross.
\newblock {Scale-Independent Inflation and Hierarchy Generation}.
\newblock {\em Phys. Lett.}, B763:174--178, 2016, 1603.05983.

\bibitem{Ferreira:2016wem}
Pedro~G. Ferreira, Christopher~T. Hill, and Graham~G. Ross.
\newblock {Weyl Current, Scale-Invariant Inflation and Planck Scale
  Generation}.
\newblock {\em Phys. Rev.}, D95(4):043507, 2017, 1610.09243.

\bibitem{Ferreira:2016kxi}
Pedro~G. Ferreira, Christopher~T. Hill, and Graham~G. Ross.
\newblock {No fifth force in a scale invariant universe}.
\newblock {\em Phys. Rev.}, D95(6):064038, 2017, 1612.03157.

\bibitem{Ferreira:2018itt}
Pedro~G. Ferreira, Christopher~T. Hill, and Graham~G. Ross.
\newblock {Inertial Spontaneous Symmetry Breaking and Quantum Scale
  Invariance}.
\newblock {\em Phys. Rev.}, D98(11):116012, 2018, 1801.07676.

\bibitem{Ferreira:2018qss}
Pedro~G. Ferreira, Christopher~T. Hill, Johannes Noller, and Graham~G. Ross.
\newblock {Inflation in a scale invariant universe}.
\newblock {\em Phys. Rev.}, D97(12):123516, 2018, 1802.06069.

\bibitem{Han:2004wt}
Tao Han and Scott Willenbrock.
\newblock {Scale of quantum gravity}.
\newblock {\em Phys. Lett.}, B616:215--220, 2005, hep-ph/0404182.

\bibitem{Bezrukov:2017dyv}
Fedor Bezrukov, Martin Pauly, and Javier Rubio.
\newblock {On the robustness of the primordial power spectrum in renormalized
  Higgs inflation}.
\newblock {\em JCAP}, 1802(02):040, 2018, 1706.05007.

\bibitem{George:2013iia}
Damien~P. George, Sander Mooij, and Marieke Postma.
\newblock {Quantum corrections in Higgs inflation: the real scalar case}.
\newblock {\em JCAP}, 1402:024, 2014, 1310.2157.

\bibitem{George:2015nza}
Damien~P. George, Sander Mooij, and Marieke Postma.
\newblock {Quantum corrections in Higgs inflation: the Standard Model case}.
\newblock {\em JCAP}, 1604(04):006, 2016, 1508.04660.

\bibitem{Bezrukov:2009db}
F.~Bezrukov and M.~Shaposhnikov.
\newblock {Standard Model Higgs boson mass from inflation: Two loop analysis}.
\newblock {\em JHEP}, 07:089, 2009, 0904.1537.

\bibitem{Fumagalli:2016lls}
Jacopo Fumagalli and Marieke Postma.
\newblock {UV (in)sensitivity of Higgs inflation}.
\newblock {\em JHEP}, 05:049, 2016, 1602.07234.

\bibitem{Branchina:2013jra}
Vincenzo Branchina and Emanuele Messina.
\newblock {Stability, Higgs Boson Mass and New Physics}.
\newblock {\em Phys. Rev. Lett.}, 111:241801, 2013, 1307.5193.

\bibitem{Branchina:2014rva}
Vincenzo Branchina, Emanuele Messina, and Marc Sher.
\newblock {Lifetime of the electroweak vacuum and sensitivity to Planck scale
  physics}.
\newblock {\em Phys. Rev.}, D91:013003, 2015, 1408.5302.

\bibitem{DiVita:2015bha}
Stefano Di~Vita and Cristiano Germani.
\newblock {Electroweak vacuum stability and inflation via nonminimal derivative
  couplings to gravity}.
\newblock {\em Phys. Rev.}, D93(4):045005, 2016, 1508.04777.

\bibitem{Xianyu:2014eba}
Zhong-Zhi Xianyu and Hong-Jian He.
\newblock {Asymptotically Safe Higgs Inflation}.
\newblock {\em JCAP}, 1410:083, 2014, 1407.6993.

\bibitem{Felder:2002jk}
Gary~N. Felder, Andrei~V. Frolov, Lev Kofman, and Andrei~D. Linde.
\newblock {Cosmology with negative potentials}.
\newblock {\em Phys. Rev.}, D66:023507, 2002, hep-th/0202017.

\bibitem{Bezrukov:2013fca}
F.~Bezrukov and D.~Gorbunov.
\newblock {Light inflaton after LHC8 and WMAP9 results}.
\newblock {\em JHEP}, 07:140, 2013, 1303.4395.

\bibitem{Akrami:2018odb}
Y.~Akrami et~al.
\newblock {Planck 2018 results. X. Constraints on inflation}.
\newblock 2018, 1807.06211.

\bibitem{GarciaBellido:2008ab}
Juan Garcia-Bellido, Daniel~G. Figueroa, and Javier Rubio.
\newblock {Preheating in the Standard Model with the Higgs-Inflaton coupled to
  gravity}.
\newblock {\em Phys. Rev.}, D79:063531, 2009, 0812.4624.

\bibitem{Repond:2016sol}
Jo~Repond and Javier Rubio.
\newblock {Combined Preheating on the lattice with applications to Higgs
  inflation}.
\newblock {\em JCAP}, 1607(07):043, 2016, 1604.08238.

\bibitem{Ema:2016dny}
Yohei Ema, Ryusuke Jinno, Kyohei Mukaida, and Kazunori Nakayama.
\newblock {Violent Preheating in Inflation with Nonminimal Coupling}.
\newblock {\em JCAP}, 1702(02):045, 2017, 1609.05209.

\bibitem{DeSimone:2008ei}
Andrea De~Simone, Mark~P. Hertzberg, and Frank Wilczek.
\newblock {Running Inflation in the Standard Model}.
\newblock {\em Phys. Lett.}, B678:1--8, 2009, 0812.4946.

\bibitem{Bezrukov:2008ej}
Fedor~L. Bezrukov, Amaury Magnin, and Mikhail Shaposhnikov.
\newblock {Standard Model Higgs boson mass from inflation}.
\newblock {\em Phys. Lett.}, B675:88--92, 2009, 0812.4950.

\bibitem{Allison:2013uaa}
Kyle Allison.
\newblock {Higgs xi-inflation for the 125-126 GeV Higgs: a two-loop analysis}.
\newblock {\em JHEP}, 02:040, 2014, 1306.6931.

\bibitem{Kannike:2015kda}
Kristjan Kannike, Antonio Racioppi, and Martti Raidal.
\newblock {Linear inflation from quartic potential}.
\newblock {\em JHEP}, 01:035, 2016, 1509.05423.

\bibitem{Ivanov:1994pa}
P.~Ivanov, P.~Naselsky, and I.~Novikov.
\newblock {Inflation and primordial black holes as dark matter}.
\newblock {\em Phys. Rev.}, D50:7173--7178, 1994.

\bibitem{Yokoyama:1998pt}
Jun'ichi Yokoyama.
\newblock {Chaotic new inflation and formation of primordial black holes}.
\newblock {\em Phys. Rev.}, D58:083510, 1998, astro-ph/9802357.

\bibitem{Hamada:2014iga}
Yuta Hamada, Hikaru Kawai, Kin-ya Oda, and Seong~Chan Park.
\newblock {Higgs Inflation is Still Alive after the Results from BICEP2}.
\newblock {\em Phys. Rev. Lett.}, 112(24):241301, 2014, 1403.5043.

\bibitem{Bezrukov:2014bra}
Fedor Bezrukov and Mikhail Shaposhnikov.
\newblock {Higgs inflation at the critical point}.
\newblock {\em Phys. Lett.}, B734:249--254, 2014, 1403.6078.

\bibitem{Garcia-Bellido:2017mdw}
Juan Garcia-Bellido and Ester Ruiz~Morales.
\newblock {Primordial black holes from single field models of inflation}.
\newblock {\em Phys. Dark Univ.}, 18:47--54, 2017, 1702.03901.

\bibitem{Ezquiaga:2017fvi}
Jose~Maria Ezquiaga, Juan Garcia-Bellido, and Ester Ruiz~Morales.
\newblock {Primordial Black Hole production in Critical Higgs Inflation}.
\newblock {\em Phys. Lett.}, B776:345--349, 2018, 1705.04861.

\bibitem{Enckell:2018kkc}
Vera-Maria Enckell, Kari Enqvist, Syksy Rasanen, and Eemeli Tomberg.
\newblock {Higgs inflation at the hilltop}.
\newblock {\em JCAP}, 1806(06):005, 2018, 1802.09299.

\bibitem{Bezrukov:2011sz}
F.~Bezrukov, D.~Gorbunov, and M.~Shaposhnikov.
\newblock {Late and early time phenomenology of Higgs-dependent cutoff}.
\newblock {\em JCAP}, 1110:001, 2011, 1106.5019.

\bibitem{Isidori:2007vm}
Gino Isidori, Vyacheslav~S. Rychkov, Alessandro Strumia, and Nikolaos Tetradis.
\newblock {Gravitational corrections to standard model vacuum decay}.
\newblock {\em Phys. Rev.}, D77:025034, 2008, 0712.0242.

\bibitem{Froggatt:1995rt}
C.~D. Froggatt and Holger~Bech Nielsen.
\newblock {Standard model criticality prediction: Top mass 173 $\pm$ 5 GeV and
  Higgs mass 135 $\pm$ 9 GeV}.
\newblock {\em Phys. Lett.}, B368:96--102, 1996, hep-ph/9511371.

\bibitem{GarciaBellido:2001ky}
Juan Garcia-Bellido, Raul Rabadan, and Frederic Zamora.
\newblock {Inflationary scenarios from branes at angles}.
\newblock {\em JHEP}, 01:036, 2002, hep-th/0112147.

\bibitem{Hamada:2013mya}
Yuta Hamada, Hikaru Kawai, and Kin-ya Oda.
\newblock {Minimal Higgs inflation}.
\newblock {\em PTEP}, 2014:023B02, 2014, 1308.6651.

\bibitem{Eichhorn:2017eht}
Astrid Eichhorn and Aaron Held.
\newblock {Viability of quantum-gravity induced ultraviolet completions for
  matter}.
\newblock {\em Phys. Rev.}, D96(8):086025, 2017, 1705.02342.

\end{thebibliography}

\end{document}